\documentclass[11pt]{article}

\usepackage[margin=1in]{geometry}
\usepackage[T1]{fontenc}
\usepackage[utf8]{inputenc}
\usepackage{amsmath,amssymb,amsthm,mathtools}
\usepackage{booktabs}
\usepackage{caption}
\usepackage{subcaption}
\usepackage{graphicx}
\usepackage{natbib}
\usepackage{setspace}
\usepackage{hyperref}
\usepackage{lmodern}

\hypersetup{
  colorlinks=true,
  linkcolor=blue,
  citecolor=blue,
  urlcolor=blue
}

\graphicspath{{./}}
\numberwithin{equation}{section}

\newtheorem{proposition}{Proposition}[section]
\newtheorem{lemma}[proposition]{Lemma}
\newtheorem{corollary}[proposition]{Corollary}
\newtheorem{assumption}[proposition]{Assumption}
\theoremstyle{definition}

\theoremstyle{remark}
\newtheorem{remark}[proposition]{Remark}

\DeclareMathOperator*{\argmax}{arg\,max}

\title{Interpreting and Countering Collusion in Deep-Learning Pricing Algorithms\thanks{I would like to thank Arhan Boyd, Yi-Chun Chen, Satoru Takahashi, and Julian Wright for helpful comments, and acknowledge support from National Natural Science Foundation of China Grant No.~W2533196.}}
\author{Soumen Banerjee\thanks{Center for Intelligence Economics, Southwestern University of Finance and Economics. E-mail: \texttt{soumen@swufe.edu.cn}}}
\date{\today}

\begin{document}

\maketitle

\begin{abstract}
Algorithmic pricing raises a question of interpretation as well as intervention: when autonomous deep-learning pricing systems sustain supracompetitive prices, what strategic pattern have they learned, and how might market institutions alter it? This paper develops an interpretable framework for studying learned collusion in repeated pricing environments. The framework embeds strategic deep learning networks in a differentiated-products Bertrand market and compresses recent price histories into finite states that record price levels, rival price movements, and movement persistence. This state representation preserves the dynamic information relevant for reward and punishment while making learned behavior economically interpretable. In the baseline environment, agents learn supracompetitive prices and exhibit a coherent collusive asymmetry: they punish rival price cuts and accommodate rival price increases. The paper then uses this framework to study an order-book mechanism that assembles temporary buyer commitments and allocates them to sellers willing to make sufficiently deep undercuts, partially insulating those undercutters from retaliatory punishment. The mechanism lowers realized prices in the main symmetric-cost design and remains effective in the main robustness exercises. Further analysis shows that this price reduction operates through the intended channel: qualifying undercuts become less exposed to subsequent punishment, reducing the continuation loss that sustains high-price states. The results show how interpretable learning frameworks can connect algorithmic pricing outcomes to economic mechanisms, and how market design can target the enforcement channel behind learned collusion.
\end{abstract}


\pagebreak
\section{Introduction}
\label{sec:introduction}

Algorithmic pricing raises a question of interpretation as well as intervention. When autonomous pricing systems repeatedly interact and sustain supracompetitive prices, policy analysis should ask not only whether prices are high, but what strategic pattern supports them and whether a market institution can alter that pattern. A growing literature shows that independent Q-learning algorithms can converge to prices above one-shot competitive benchmarks in repeated Bertrand environments \citep{calvano2020artificial,klein2021autonomous}. Empirical evidence also suggests that algorithmic pricing can materially change prices in real markets \citep{brown2023competition,clark2023algorithmic}. Several papers have also studied potential policy interventions involving platform-driven demand steering \citep{johnson2020platform}, and audit protocols \citep{hartline2024regulation, hartline2025refined}. These findings have made algorithmic collusion a concrete object of economic analysis.

The central difficulty is that high average prices are not, by themselves, a complete economic object. For policy purposes, the relevant question is whether the learned policy has the dynamic structure associated with collusive enforcement. In a repeated pricing game, theory predicts that supracompetitive prices are sustained by dynamic incentives: firms refrain from undercutting because price cuts trigger punishment related losses, while movements toward higher prices are accommodated. However, recent evidence \citep{epivent2024rewardpunishment} shows that the canonical Q-learning pricing algorithm punishes price rises as well, behavior that complicates the interpretation of the learned policy as a plausible commercial pricing strategy, especially in environments where price increases may reflect cost or demand conditions. A useful framework for evaluating policy responses to algorithmic collusion should therefore generate supracompetitive prices and exhibit a recognizable reward-punishment logic.

This paper develops such an interpretable framework for studying learned collusion in deep-learning pricing algorithms. The framework uses strategic deep Q-network learners in a repeated differentiated-products Bertrand environment.\footnote{\cite{clark2023algorithmic} note in their study of the German retail gasoline market that the pricing software used, Pricecast Fuel, relies on deep-learning neural network models. This is a natural modeling choice because deep learning allows for more flexible function approximation than the tabular Q-learning algorithms commonly studied in the literature.} The state representation is motivated by the repeated-games literature on finite automata, bounded memory, and strategic complexity \citep{rubinstein1986finiteautomata,abreu1988finiteautomata,kalai1988finite,sabourian1998boundedmemory,barlo2009onememory,barlo2016boundedmemory}. Rather than feeding the learner a raw price history, the framework compresses recent play into a compact strategic state: own price level, rival price level, rival movement type, and rival movement duration. The price-level components describe the current pricing environment, while the rival-movement components record whether the rival has recently cut, held, or raised price and whether that pattern has persisted.

This representation gives the learner a compact economic language for conditioning on strategically relevant histories. It preserves the directional and persistence information central to reward-punishment accounts of collusion, while avoiding an opaque dependence on long raw histories. The action space is similarly interpretable, consisting of relative-price moves such as large cuts, moderate cuts, matching, and upward following.\footnote{This structure aggregates strategically similar but numerically distinct histories, improving the sample efficiency of the learning process. It also permits the learner to condition on longer effective histories, up to 8 periods, than the short histories typically used in tabular Q-learning benchmarks.} Together, the state and action representations simplify the learning problem while making the resulting policy behavior economically interpretable.

In the baseline environment, the agents learn an interpretable form of supracompetitive pricing. In the main symmetric-cost design, the final-policy greedy rollout price is 1.6461, about 11.75 percent above the one-shot Nash price of 1.4729.\footnote{In machine learning, the term ``greedy'' is used to refer to a final policy rollout, i.e. when the learner plays the learnt strategy with no experimentation.} More importantly, rise/drop validation exercises show that the trained learner responds asymmetrically to rival price movements: it lowers prices after rival price cuts and raises prices after rival price increases. The framework therefore does not merely generate high average prices; it generates high prices supported by behavior consistent with the reward-punishment logic predicted by repeated-game theory.

The paper then asks whether a market institution can weaken this enforcement logic. The intervention is an \emph{order-book} mechanism: an operator solicits conditional buyer commitments at prices below a recent reference price, aggregates them into a temporary demand block, and offers that block to sellers. A seller qualifies by agreeing to supply the committed demand at a sufficiently low contract price; if multiple sellers qualify, the lowest qualifying seller fills the book. The winning seller receives protected committed demand for a finite duration, while residual demand continues to be served through the ordinary spot market. The operator may be a marketplace platform or an independent buyer-side aggregator that earns a small spread on the commitments.

The mechanism targets the enforcement margin that sustains supracompetitive pricing. Without the mechanism, a seller considering an undercut trades off the current gain from a lower price against the continuation loss generated by future punishment. By protecting part of the deviator's demand after a qualifying undercut, the order book mechanism reduces that continuation loss and raises the net value of deviation. The design is therefore most natural in repeat-purchase or procurement markets where temporary buyer commitment is feasible, including retail gasoline \citep{clark2023algorithmic} and generic pharmaceuticals.

The computational design keeps the intervention separate from the learner's information structure. Order-book variables are excluded from the agents' state representation and action space. The baseline environment and the environment with the mechanism use the same strategic state, six-action output space, network architecture, training horizon, and exploration schedule. The intervention changes realized allocation and payoffs while leaving the agents' observed state and available actions fixed. This design makes the price difference interpretable as a payoff-side consequence of market design rather than as a change in what the algorithm observes or can choose.

The main computational analysis evaluates the mechanism at two levels. First, the main evaluation trains policies from scratch in the baseline environment and the environment with the mechanism. In the symmetric-cost setting, the mechanism lowers final-policy realized transaction prices. Relative to the one-shot Nash price, the policy learned in the baseline environment is 11.75 percent above Nash, while the policy learned in the environment with the mechanism is 6.23 percent above Nash. Equivalently, the mechanism closes about 47.0 percent of the baseline collusion gap.

Second, the paper uses same-state forced-deviation diagnostics to isolate the channel behind this price difference. Starting from common high-price histories, one path holds both sellers at the high price, while the deviation path forces one seller to make an undercut which qualifies them for the order book, and then releases both sellers back to greedy play. This exercise decomposes the deviation incentive into the period-0 payoff from the forced cut, the continuation loss from subsequent punishment, and the resulting net deviation value. It therefore separates the dynamic punishment channel from the ordinary one-period price-cutting incentive.

The forced-deviation results align closely with the theory. Relative to the baseline environment, the order-book mechanism reduces the continuation loss from punishment and increases the net value of a qualifying deviation:  the same qualifying cut is unprofitable in the baseline environment but profitable in the environment with the mechanism, because the deviator is protected during the punishment phase. The price reduction is therefore linked to a specific economic channel: qualifying undercuts become less exposed to future retaliation. Additional high-start rollouts from common high-price histories support this interpretation, showing that policies learned in the environment with the mechanism leave high-price states more quickly and end at lower prices.

A heterogeneous-cost extension with different seller costs shows that the mechanism remains effective away from exact symmetry, yielding lower realized rollout prices and closing about 35.6 percent of the heterogeneous-cost collusion gap.

The paper makes three contributions. First, it develops an interpretable framework for studying learned collusion in deep-learning pricing algorithms. The framework uses strategic DQN learners but organizes the learning problem around economically meaningful state and action variables. Price histories are compressed into a finite strategic state that captures price levels, rival movement direction, and movement persistence, while the action space consists of interpretable relative-price moves. This structure makes the learned policy economically interpretable: in the baseline environment, perturbation tests show that the learner punishes rival price cuts and accommodates rival price increases, so supracompetitive prices are not merely a computational finding; they are supported by the reward-punishment pattern predicted by economic theory.

Second, the paper introduces an order-book mechanism as a specific instance of a market-design intervention against learned collusive pricing. The mechanism uses temporary buyer commitment to protect qualifying undercuts and thereby weaken the continuation-value logic that sustains supracompetitive prices. It can be run by a platform or by an independent buyer-side aggregator. Its distinction from platform-driven demand steering, such as dynamic PDP \citep{johnson2020platform}, is that it works through commitment rather than exclusivity. The order book protects only a temporary, partial block of committed demand, while residual demand remains available through ordinary spot-market competition. It therefore weakens the punishment channel that sustains collusion without making market fulfillment depend on a single favored seller. Its distinction from ex post audit protocols \citep{hartline2024regulation,hartline2025refined} is that it is not a certification rule for past behavior, but an ex ante change to the payoff environment in which pricing algorithms learn. It therefore does not require the operator to classify sellers' pricing algorithms or certify past pricing behavior.

Third, consistent with the interpretability-forward approach taken here, the paper connects the reduced-form price reduction to the enforcement channel isolated by the framework. Because the intervention changes realized allocations and payoffs while leaving the agents' state representation and action space fixed, same-state diagnostics can isolate how the order book changes deviation incentives. The forced-deviation tests show that the mechanism reduces the continuation loss from punishment and increases the net value of a qualifying deviation by protecting the deviator during the punishment phase. The rollout, high-start, and heterogeneous-cost exercises then show how this incentive channel maps into learned pricing outcomes. The central contribution is therefore not only that the order book lowers prices, but that the framework explains why: the intervention makes undercutting less vulnerable to the retaliatory dynamics that sustain learned collusion.

The rest of the paper proceeds as follows. Section \ref{sec:benchmark} presents the strategic DQN framework and validates its learned behavior. Section \ref{sec:model} develops the repeated-pricing incentive logic behind the order-book mechanism. Section \ref{sec:experiment} compares the baseline environment and the environment with the mechanism, reports mechanism diagnostics, and studies extensions. Section \ref{sec:literature} reviews the related literature on algorithmic collusion, bounded-memory representations, deep reinforcement learning, and interventions. Section \ref{sec:conclusion} concludes. %Section \ref{sec:discussion} discusses interpretation, scope conditions, and limitations. 

\section{Strategic DQN Framework and Behavioral Validation}
\label{sec:benchmark}

\subsection{Strategic DQN learner}
\label{subsec:dqn}

The framework embeds a repeated Bertrand pricing game played by two independent reinforcement-learning sellers. Each seller uses a deep Q-network (DQN), which can be understood as a rule that assigns a continuation value to each available pricing move in each observed state. In every period, a seller observes a finite strategic state, chooses a pricing move, earns its current profit, and then updates its value estimates from the realized transition. The important modeling choice is therefore not the neural-network architecture itself, but the economic language in which the learner is allowed to condition its behavior.

The framework operationalizes the bounded-memory idea from repeated-game theory by mapping recent public price histories into a small strategic state. The simulator keeps an eight-period price history, but the learner is not given that raw sequence. Instead, the history is compressed into four economically interpretable variables: the seller's own current price level, the rival's current price level, the direction of the rival's recent movement, and the persistence of that movement. The movement variable distinguishes sharp drops, mild drops, holds, mild rises, and sharp rises; the duration variable distinguishes short-lived, medium-lived, and persistent movements. This produces 375 possible strategic states. The representation is not claimed to be a sufficient statistic for all repeated-pricing behavior. It is a deliberately finite, theory-guided state abstraction whose adequacy is assessed by behavioral diagnostics.

The action space is also chosen to be economically legible. Instead of choosing an arbitrary continuous price, the learner chooses one of six relative moves around the rival's current price: a large cut, a qualifying three-grid-step cut, a one-step tactical undercut, a match, a one-step upward follow, or a three-step upward follow. These moves span the main responses relevant for collusive pricing: punish a rival by cutting, match the rival, undercut slightly, or follow the rival upward. The ``qualifying cut'' label reflects the order-book mechanism's three-step qualification threshold, but the action is available with the same definition whether or not the mechanism is active.

The learner remains blind to the order-book mechanism. In both the baseline environment and the environment with the mechanism, the learner observes the same strategic state and chooses from the same six actions. No order-book variable is added to the state, and no special action is added to the action set. This restriction is central to the interpretation of the evaluation: the mechanism changes the payoff consequences of pricing decisions, not the information available to the pricing algorithm.

\subsection{Training and evaluation logic}
\label{subsec:state_actions_training}

Training is from scratch. In the baseline environment, the order-book mechanism is inactive. In the environment with the mechanism, otherwise identical sellers repeatedly interact with the order-book institution active from the beginning. In each case, 100 different runs with different random seeds (1-100) are trained. The state representation, action set, learning algorithm, training horizon, and exploration schedule are held fixed across the two environments. Thus the comparison asks whether the same class of pricing algorithms learns a different pricing policy when the market institution changes.

During training, sellers sometimes experiment rather than choosing their currently highest-valued action. This exploration is useful for learning, but it means that training-period prices combine the emerging policy with deliberate random experimentation. For that reason, the paper evaluates trained agents using final-policy greedy rollouts, i.e. rollouts which do not feature any experimentation. A greedy rollout turns off exploration and asks what prices the learned policy chooses when deployed. The figures and tables below therefore report greedy-rollout realized transaction prices unless otherwise stated. Appendix \ref{app:implementation} gives the implementation recipe: the exact state encoder, action-to-price map, value-update rule, exploration schedule, reset rule, and evaluation starts.

\subsection{Baseline Supracompetitive Pricing}
\label{subsec:baseline_result}

The baseline environment uses the same two-firm differentiated-products Bertrand-logit environment as \citet{calvano2020artificial}. Two symmetric sellers have common marginal cost \(c=1.0\), demand follows the logit specification with \(A=2.0\), outside option \(A_0=0.0\), and \(\mu=0.25\), and prices are chosen from a 15-point grid between 1.00 and 2.10. In this economic environment, the one-shot symmetric Nash price is 1.4729 and the symmetric monopoly price is 1.9250.

% User comment: Report percentages for the baseline environment and explicitly show the collusion-gap calculation.
% Fix: Added percent-over-Nash, percent-below-monopoly, and the baseline collusion-gap arithmetic.

\begin{figure}[t]
	\centering
	\includegraphics[width=0.78\textwidth]{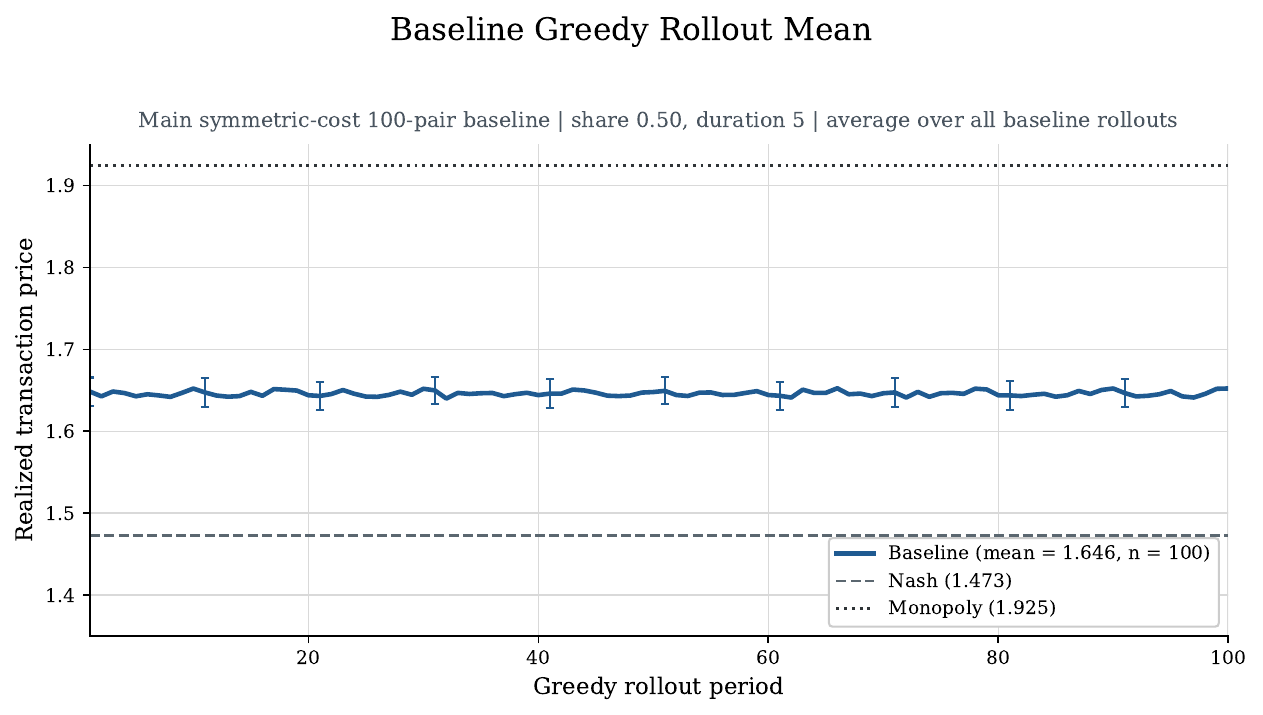}
	\caption{Baseline greedy rollout. The figure reports the baseline symmetric-cost final-policy rollout mean, averaged over 100 baseline runs. The horizontal lines represent the one-shot Nash and monopoly prices.}
	\label{fig:baseline_without}
\end{figure}

Figure \ref{fig:baseline_without} shows the baseline environment. It reports the mean realized transaction price from the baseline final-policy greedy rollout in the main symmetric-cost 100-pair batch. The mean realized price is 1.6461. This is about 11.75 percent above the symmetric Nash price of 1.4729, since \((1.6461-1.4729)/1.4729\approx 0.1175\), and about 14.49 percent below the monopoly price of 1.9250. The baseline collusion gap is \(1.6461-1.4729=0.1732\), or about 38.3 percent of the Nash-to-monopoly interval \(1.9250-1.4729\). The baseline environment is supracompetitive enough to make a mechanism test meaningful and leaves room for economically plausible adjustment.

The figure reports the baseline final-policy object. The policy learned in the baseline environment is the reference point against which the order-book institution is evaluated. It establishes that the framework generates the type of supracompetitive pricing that a policy intervention should seek to weaken.

\subsection{Behavioral validation: punishment of cuts and accommodation of rises}
\label{subsec:behavioral_validation}

A framework intended for policy analysis should exhibit a plausible strategic logic. The validation exercise perturbs the rival's price during a baseline rollout and measures the focal agent's response. The key question is directional. If the rival cuts price, a collusive learner should respond by cutting price. If the rival raises price, a collusive learner should generally accommodate the increase.

% User comment: Use the five-period-held rise/drop panels, avoid "main draft" language, and convert response magnitudes into grid-step units.
% Fix: Switched Figure 2 to the hold-five panels, wrote "I report," and used the aggregate JSON to report shock-end responses in price units and grid-step units.

\begin{figure}[t]
	\centering
	\begin{subfigure}{0.48\textwidth}
		\centering
		\includegraphics[width=\textwidth]{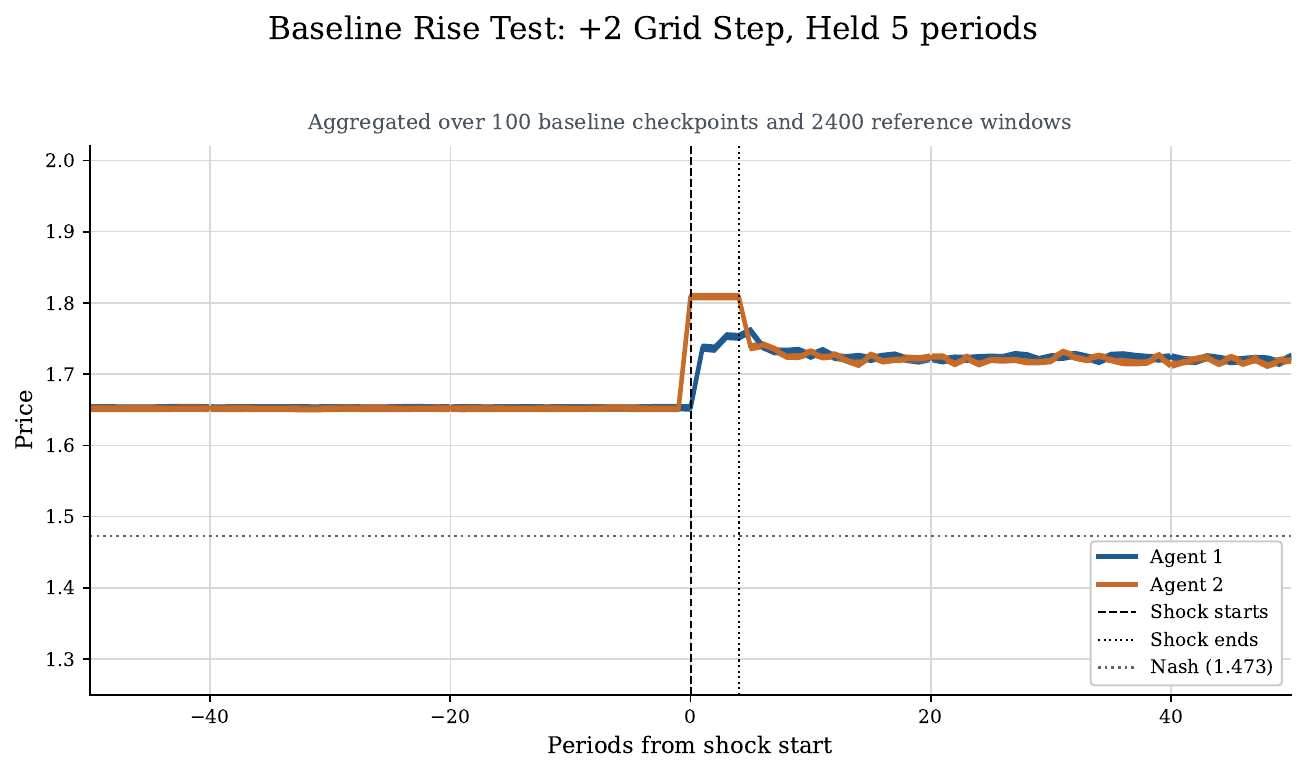}
		\caption{Rise, +2 steps, held 5}
	\end{subfigure}
	\hfill
	\begin{subfigure}{0.48\textwidth}
		\centering
		\includegraphics[width=\textwidth]{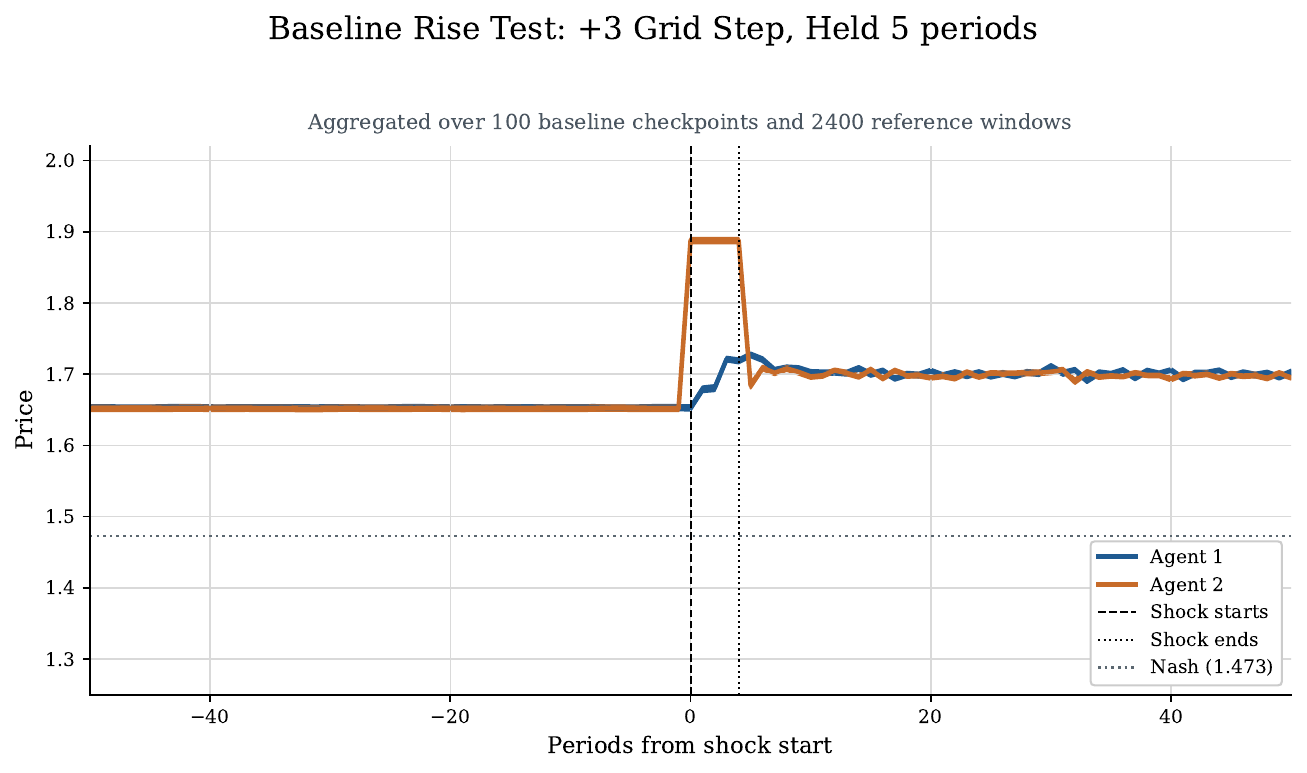}
		\caption{Rise, +3 steps, held 5}
	\end{subfigure}
	
	\vspace{0.5em}
	\begin{subfigure}{0.48\textwidth}
		\centering
		\includegraphics[width=\textwidth]{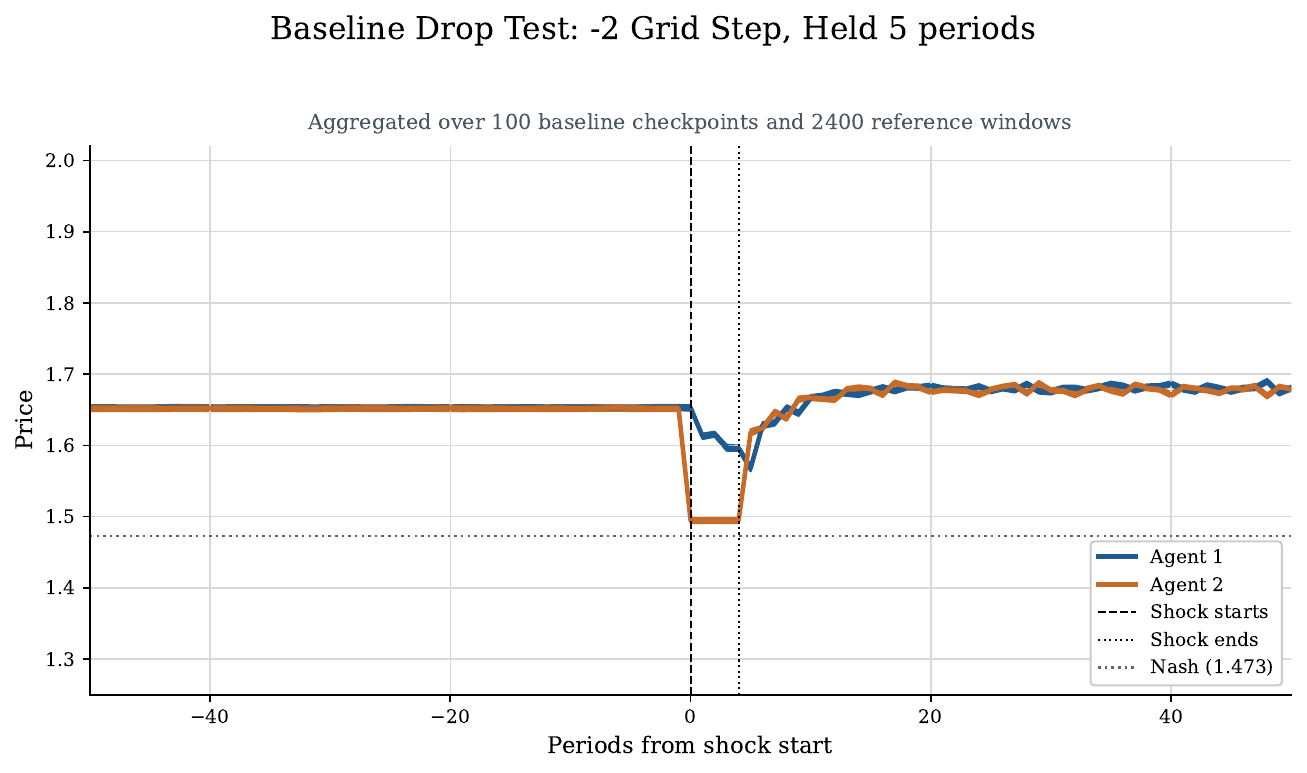}
		\caption{Drop, -2 steps, held 5}
	\end{subfigure}
	\hfill
	\begin{subfigure}{0.48\textwidth}
		\centering
		\includegraphics[width=\textwidth]{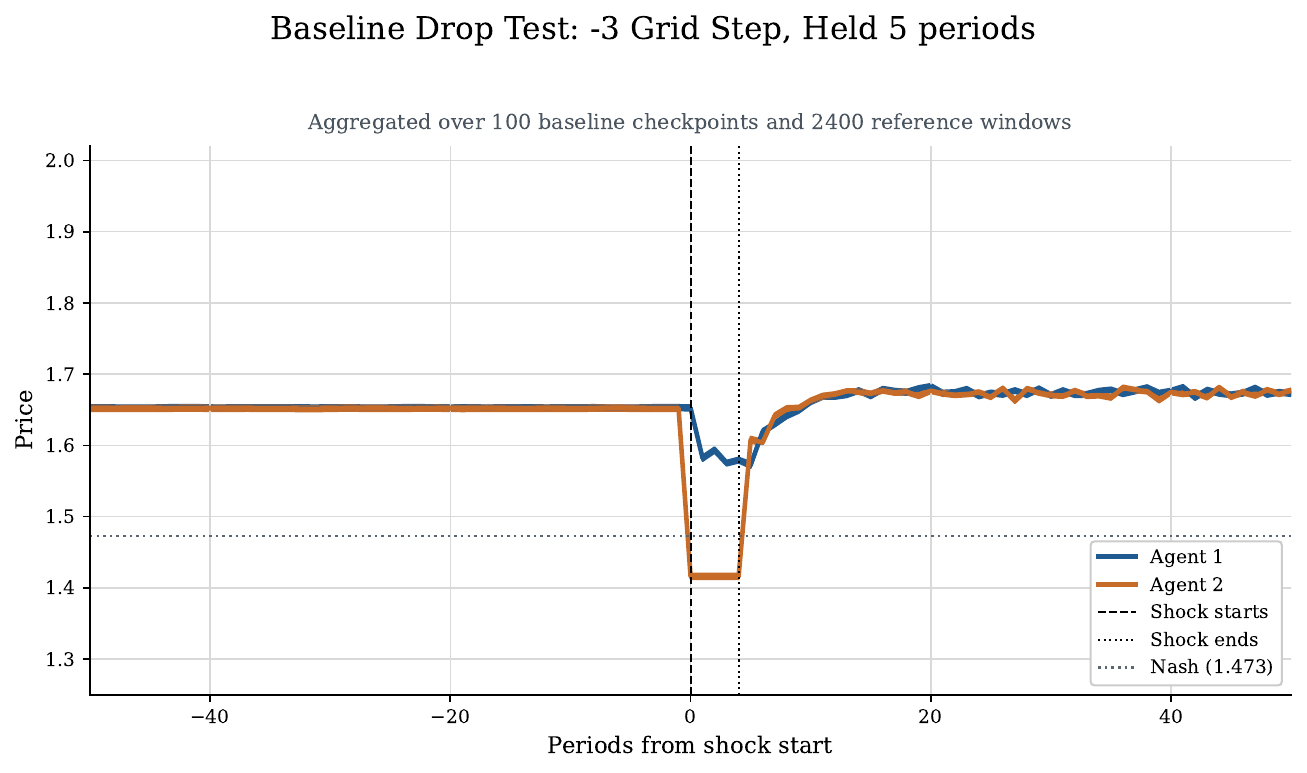}
		\caption{Drop, -3 steps, held 5}
	\end{subfigure}
	\caption{Behavioral validation through rise and drop perturbations. The figure reports held-shock two- and three-step perturbation tests for the baseline environment, with each perturbation held for five periods. The focal learner generally follows rival price increases and cuts prices after rival price drops.}
	\label{fig:rise_drop_validation}
\end{figure}

Figure \ref{fig:rise_drop_validation} reports the held-shock rise/drop validation exercises, focusing on two- and three-grid-step rival perturbations held for five periods. The held-shock paths make the directional response especially easy to see. The tests aggregate across 100 baseline checkpoints and 2,400 reference windows per specification. The price grid has step size about 0.0786. By the end of a held two-step rival price increase, the focal agent raises its price by about 0.099, or about 1.27 grid steps. By the end of a held three-step increase, the focal response remains positive, about 0.066, or about 0.84 grid steps. After a held two-step rival price drop, the focal agent lowers its price by about 0.057, or about 0.73 grid steps; after a held three-step drop, it lowers price by about 0.074, or about 0.94 grid steps. The magnitude varies with the shock size. The central fact is directional: upward deviations are generally accommodated and downward deviations are punished.

This validation is a behavioral diagnostic. It shows that the baseline environment has the directional asymmetry one would expect from a policy that protects high prices against undercutting and incorporates price increases into the path. That makes it a more credible test bed for evaluating a market-design intervention.

\section{Model and Mechanism}
\label{sec:model}

% User comment: Soften overly aggressive language in the theory section where possible.
% Fix: Replaced stronger "destabilizes collusion" language with "weakens collusive enforcement" while keeping the formal incentive result intact.
This section provides a deliberately stylized rational model for the order-book mechanism. The analysis isolates the economic channel using a streamlined representation of details such as the trailing-window reference price, the grid-step qualification rule, and the architecture of the learner. In the repeated pricing environment studied here, supracompetitive pricing can be sustained only if a detected undercut lowers the deviator's continuation value enough to offset the ordinary one-period deviation gain. The order-book mechanism weakens collusive enforcement by reducing exactly that continuation-value loss.

\subsection{Environment}
\label{subsec:theory_environment}

% User comment: Note that marginal cost is identical here and that asymmetric costs are studied later.
% Fix: Described \(c\) as common marginal cost and cross-referenced the asymmetric-cost extension.
Consider two sellers \(i\in\{1,2\}\) that interact in periods \(t=0,1,2,\dots\). In each period seller \(i\) chooses a price \(p_{it}\in\mathcal P\), where \(\mathcal P\subset[c,\bar p]\) is a finite grid and \(c\) is the common marginal cost. Seller-specific costs are introduced in the asymmetric-cost extension in Subsection \ref{subsec:theory_asymmetric}. Let
\[
\pi_i(p_i,p_j)
\]
% User comment: Mention that the demand environment includes Bertrand-logit settings such as Johnson and Calvano.
% Fix: Added citations to Calvano and Johnson in the demand-system sentence.
denote seller \(i\)'s one-period profit when the price profile is \((p_i,p_j)\). The results below apply across standard demand systems, including Bertrand-logit environments of the kind used in \citet{calvano2020artificial} and related platform-design work such as \citet{johnson2020platform}.

\begin{assumption}[Stage game]
\label{ass:stage}
The one-shot pricing game has a unique symmetric Nash price \(p^N\in\mathcal P\).
\end{assumption}

Fix a candidate supracompetitive collusive price \(p^C>p^N\), and let
\[
\pi^C := \pi_i(p^C,p^C)
\]
denote the per-period collusive profit of each seller.

An operator runs a temporary customer-commitment mechanism. The operator may be the marketplace platform itself or an independent buyer-side aggregator. The operator first solicits conditional buyer commitments at candidate prices below a reference price and thereby creates an order book, then offers that order book to sellers. In the reduced form used below, a seller buys the book by agreeing to supply the pre-assembled demand block at a sufficiently low qualifying contract price. Let \(r\) denote a reference price and let \(\kappa>0\) denote the minimum discount required to qualify for the order book. A candidate supply price \(q\in\mathcal P\) is \emph{qualifying} if
\[
q \le r-\kappa.
\]
Along a stationary collusive path it is natural to set \(r=p^C\), so qualifying undercuts satisfy \(q\le p^C-\kappa\). Assume the set of qualifying prices is nonempty, and let
\[
q^\star \in \argmax_{q\in\mathcal P:\, q\le p^C-\kappa}\pi_i(q,p^C),
\qquad
\pi^D := \pi_i(q^\star,p^C).
\]
Thus \(\pi^D\) is the one-period profit from the best qualifying undercut against the collusive price.

We restrict attention to a transparent class of equilibria: symmetric public perfect equilibria with a collusive phase and a finite punishment phase. In the collusive phase both sellers charge \(p^C\). If seller \(i\) is detected undercutting in the collusive phase, then beginning next period the game moves to a publicly observed punishment path directed at that deviator; after \(T<\infty\) punishment periods, play returns to the collusive phase.

Let \(W^C\) denote the deviator's continuation value from next period onward if no deviation is observed today, and let \(W^D\) denote the same seller's continuation value from next period onward if a detected qualifying undercut is observed today. The theory below focuses on the collusive-phase incentive constraint. That constraint is necessary for sustaining any supracompetitive path, and it is exactly the constraint that the order-book mechanism is designed to weaken.

For clarity, the main comparative-statics statements below assume that the order-book duration \(L\) is weakly below the punishment length \(T\). If the book lasts longer than the punishment phase, one replaces the punishment-profit sequence in the protection term by the relevant post-deviation profit sequence; the argument is unchanged.

\subsection{Repeated pricing without intervention}
\label{subsec:theory_without}

\begin{lemma}[Punishment necessity / continuation-value condition]
\label{lem:cv}
For any admissible supracompetitive equilibrium, a detected qualifying undercut can be deterred only if
\begin{equation}
\label{eq:cv}
\pi^D-\pi^C \le \delta\bigl(W^C-W^D\bigr),
\end{equation}
where \(\delta\in(0,1)\) is the common discount factor.
Equivalently, any supracompetitive repeated-price equilibrium must make a detected undercut unprofitable by lowering the deviator's continuation value.
\end{lemma}

\begin{proof}
If the seller complies today, its value is
\[
V^C=\pi^C+\delta W^C.
\]
If it deviates today to the best qualifying undercut \(q^\star\), its value is
\[
V^D=\pi^D+\delta W^D.
\]
Sequential rationality requires \(V^C\ge V^D\). Rearranging yields \eqref{eq:cv}.
\end{proof}

\begin{corollary}[Finite-punishment form]
\label{cor:finite_punishment}
Suppose a detected undercut triggers \(T\) punishment periods during which the deviator's profits are
\[
\pi_0^P,\pi_1^P,\dots,\pi_{T-1}^P,
\]
after which the game returns to the collusive phase. Then \eqref{eq:cv} becomes
\begin{equation}
\label{eq:finite_ic}
\pi^D-\pi^C
\le
\delta\sum_{s=0}^{T-1}\delta^s\bigl(\pi^C-\pi_s^P\bigr).
\end{equation}
\end{corollary}

\begin{proof}
Under the assumed structure,
\[
W^C=\sum_{s=0}^{T-1}\delta^s\pi^C+\delta^T V^C
\quad\text{and}\quad
W^D=\sum_{s=0}^{T-1}\delta^s\pi_s^P+\delta^T V^C,
\]
where \(V^C\) is the value of returning to the collusive phase. Hence
\[
W^C-W^D
=
\sum_{s=0}^{T-1}\delta^s\bigl(\pi^C-\pi_s^P\bigr).
\]
Substituting this into \eqref{eq:cv} yields \eqref{eq:finite_ic}.
\end{proof}

\begin{remark}[Interpretation]
Lemma \ref{lem:cv} formalizes the notion that supracompetitive pricing must be enforced by punishment. In general repeated-game language, the relevant enforcement object is the continuation-value drop \(W^C-W^D\) following a detected undercut. The finite-punishment representation in Corollary \ref{cor:finite_punishment} makes that continuation-value loss especially transparent.
\end{remark}

\subsection{Order-book protection and protected deviation value}
\label{subsec:theory_protection}

The mechanism can be understood most simply through a numerical retail-gasoline
example. Suppose nearby stations have recently charged a reference price of
\(r=1.80\), and suppose the operator defines a discount of at least three grid
steps, approximately \(\kappa=0.24\), as a qualifying offer. Consider a
candidate contract price \(q=1.56\), so \(q\le r-\kappa\). Buyers who are
willing to commit some future purchases at that price create an order book. If,
for example, they commit a fraction \(\lambda=0.50\) of normalized demand for
\(L=5\) periods, the result is a temporary order book at contract price
\(q=1.56\). The operator then offers this order book to sellers, and a station
willing to supply at that price can buy it. The remaining half of demand
continues to be served in the ordinary spot market.

This timing is important. The order book is not an exogenous transfer to the
deviating seller, and it is not something created only after the seller has
already deviated. It is first assembled from buyer commitments at a candidate
contract price and only then offered to sellers. A low enough price therefore
has two consequences. First, it may attract ordinary spot demand in the current
period. Second, it can give the seller access to a protected block of committed
demand that remains in place during later punishment periods. The second consequence
is the distinctive feature of the mechanism.

The key economic point is that the order book changes what happens after a
deviation is detected. Without the mechanism, a seller that undercuts the
collusive price can be punished in later periods by being forced into a low
spot-market payoff. With the mechanism, the same seller remains partly exposed
to the spot market, but a protected block of demand continues to transact at
the previously agreed contract price. For example, if marginal cost is \(c=1\)
and the deviator's ordinary punishment profit in a period is \(\pi_s^P=0.20\),
then the protected block alone yields margin \(q-c=0.56\). With
\(\lambda=0.50\), the protected-period profit in the reduced form below is
\[
(1-\lambda)\pi_s^P+\lambda(q-c)
=
0.5(0.20)+0.5(0.56)
=
0.38.
\]
Thus the order book raises the deviator's punishment-period payoff from
\(0.20\) to \(0.38\) in that protected period. The exact numbers are only
illustrative; the mechanism works by making punishment less severe after a
qualifying undercut.

Formally, fix a candidate price \(q\). Qualification means that the candidate
supply price is sufficiently below the relevant reference price, \(q\le r-\kappa\).
The timing of the mechanism is:
\begin{enumerate}
\item The operator offers buyers a forward purchase opportunity at price \(q\).
\item The resulting commitments form an order book that allocates a fraction
\(\lambda\in(0,1)\) of demand for \(L\) periods at contract price \(q\).
\item The operator then offers this order book to sellers.
\item A seller that agrees to supply at the qualifying price \(q\)
buys the book; if several sellers are willing to do so, the lowest-price
qualifying seller gets to fill it.
\end{enumerate}

In the reduced form analyzed below, these four steps are summarized by representing
the qualifying seller as taking up a temporary order book of share \(\lambda\)
and duration \(L\) at contract price \(q\), while the remaining demand
continues to be allocated through the spot market.

Let \(W^{D,OB}(q,\lambda,L)\) denote the deviator's continuation value from
next period onward after a detected qualifying undercut when the order book is
active, and let \(W^D\) denote the corresponding continuation value without the
order book. Define the \emph{protected deviation value} by
\begin{equation}
	\label{eq:defB}
	B(q,\lambda,L)
	:=
	\delta\Bigl(W^{D,OB}(q,\lambda,L)-W^D\Bigr).
\end{equation}
Thus \(B(q,\lambda,L)\) is the extra discounted continuation value created by
the order book, measured at the date of deviation. To connect that
continuation-value object to period profits, it is useful to work with the
following reduced form.

If the deviator would otherwise earn punishment profit \(\pi_s^P\) in
punishment period \(s\), then with order-book protection it earns
\begin{equation}
\label{eq:protected_profit}
\pi_s^{P,OB}(q,\lambda)
=
(1-\lambda)\pi_s^P+\lambda(q-c),
\qquad s=0,1,\dots,L-1,
\end{equation}
and \(\pi_s^{P,OB}(q,\lambda)=\pi_s^P\) for \(s\ge L\). Equation \eqref{eq:protected_profit} should be read as a reduced form in which a fraction \(\lambda\) of unit demand is carved out from the spot market and transacted at the contract price \(q\). The remaining fraction \(1-\lambda\) continues to generate the ordinary punishment profit.

The fixed-contract formulation takes the contract price and marginal cost as fixed over the life of the book; indexed contracts that pass through observable input-cost movements are discussed in Remark~\ref{rem:indexed_contracts}.

% User comment: Implement the transcript discussion on fixed-demand protection, possible demand expansion, and conservative protected-profit monotonicity.
% Fix: Added a conservative-reduced-form paragraph and stated the protected-profit dominance condition under which richer demand-expansion models weakly strengthen the mechanism.
The fixed-demand specification is a conservative reduced form. It intentionally holds the protected quantity fixed at \(\lambda\), so a lower contract price affects protected profit through the per-unit margin \(q-c\) rather than through an endogenous expansion of the committed block. A richer model could let the protected quantity respond to the contract price. Let
\[
\phi_0(q,\lambda,\pi_s^P):=(1-\lambda)\pi_s^P+\lambda(q-c)
\]
denote the fixed-demand protected-profit benchmark in \eqref{eq:protected_profit}, and let \(\phi(q,\lambda,\pi_s^P)\) denote any richer protected-profit technology, including one with buyer-side demand expansion. If
\[
\phi(q,\lambda,\pi_s^P)\ge \phi_0(q,\lambda,\pi_s^P)
\]
in every protected period, then the associated protection value is weakly larger than in the fixed-demand benchmark. Thus any demand-expansion channel that raises protected profit period by period weakly tightens the incentive constraint for collusive pricing. The condition is stated in profit terms because larger quantity alone is not sufficient: a lower contract price also lowers the per-unit margin.

\begin{lemma}[Order-book protection value]
	\label{lem:protection}
	Under \eqref{eq:protected_profit},
	\begin{equation}
		\label{eq:B_formula}
		B(q,\lambda,L)
		=
		\delta\sum_{s=0}^{L-1}\delta^s\Bigl(\pi_s^{P,OB}(q,\lambda)-\pi_s^P\Bigr)
		=
		\delta\lambda\sum_{s=0}^{L-1}\delta^s\Bigl[(q-c)-\pi_s^P\Bigr].
	\end{equation}
	Hence:
	\begin{enumerate}
		\item[(i)] if \(q-c\ge \pi_s^P\) for every \(s=0,\dots,L-1\), with strict inequality for at least one such \(s\), then \(B(q,\lambda,L)>0\);
		\item[(ii)] if \(q-c\ge \pi_s^P\) for all \(s=0,\dots,L-1\), then \(B(q,\lambda,L)\) is weakly increasing in \(\lambda\), and strictly increasing if the inequality is strict for at least one such \(s\) and \(\lambda>0\);
		\item[(iii)] if \(q-c\ge \pi_s^P\) in each newly added protected period, then \(B(q,\lambda,L)\) is weakly increasing in \(L\), with strict increase whenever the newly added period satisfies \(q-c>\pi_L^P\).
	\end{enumerate}
\end{lemma}

\begin{proof}
	The first equality is just the definition of \(B\) specialized to the finite-punishment representation. The second follows from \eqref{eq:protected_profit}:
	\[
	\pi_s^{P,OB}(q,\lambda)-\pi_s^P
	=
	(1-\lambda)\pi_s^P+\lambda(q-c)-\pi_s^P
	=
	\lambda\bigl[(q-c)-\pi_s^P\bigr].
	\]
	Substituting into \eqref{eq:defB} yields \eqref{eq:B_formula}.
	
	For part (i), \(B(q,\lambda,L)\) is a discounted sum of weakly nonnegative terms, at least one of which is strictly positive. Hence the sum is strictly positive.
	
	For part (ii), differentiating \eqref{eq:B_formula} with respect to \(\lambda\) gives
	\[
	\frac{\partial B(q,\lambda,L)}{\partial\lambda}
	=
	\delta\sum_{s=0}^{L-1}\delta^s\bigl[(q-c)-\pi_s^P\bigr].
	\]
	If \(q-c\ge \pi_s^P\) for all \(s=0,\dots,L-1\), every term in the sum is non‑negative, so the derivative is non‑negative and \(B\) is weakly increasing in \(\lambda\). If the inequality is strict for at least one period, the sum is strictly positive and \(B\) is strictly increasing in \(\lambda\) (provided \(\delta>0\)).
	
	For part (iii),
	\[
	B(q,\lambda,L+1)-B(q,\lambda,L)
	=
	\delta^{L+1}\lambda\bigl[(q-c)-\pi_L^P\bigr].
	\]
	This is weakly nonnegative whenever the newly added protected period satisfies \(q-c\ge \pi_L^P\), and strictly positive if the inequality is strict.
\end{proof}

% User comment: Implement the transcript discussion by making the conservative protected-profit comparison explicit.
% Fix: Replaced the shorter generality remark with a protected-profit dominance statement for \(B_\phi\).
\begin{remark}[More general protected-profit technologies]
The argument extends beyond the linear specification in \eqref{eq:protected_profit}. For any protected-profit function \(\phi(q,\lambda,\pi_s^P)\), define
\[
B_\phi(q,\lambda,L)
:=
\delta\sum_{s=0}^{L-1}\delta^s\bigl(\phi(q,\lambda,\pi_s^P)-\pi_s^P\bigr).
\]
If \(\phi(q,\lambda,\pi_s^P)\ge \phi_0(q,\lambda,\pi_s^P)\) in every protected period, then \(B_\phi(q,\lambda,L)\ge B(q,\lambda,L)\), with strict inequality if the protected-profit dominance is strict in at least one period. A sufficient condition in a quantity-expansion model is \(m(q,\lambda)\ge\lambda\) and \(q-c\ge \pi_s^P\) in each protected period, where
\[
\phi(q,\lambda,\pi_s^P)
=
(1-m(q,\lambda))\pi_s^P+m(q,\lambda)(q-c).
\]
The fixed-demand benchmark therefore gives a transparent lower bound on the protection value whenever richer demand responses weakly raise protected profit.
\end{remark}

\begin{remark}[Take-up-period protection]
If the institutional design lets the seller serve committed demand in the same
period in which it takes up the book, one can either absorb that same-period
protected-demand increment into \(\pi^D\), or equivalently start the protection
term as beginning one period earlier. This is only an accounting convention:
the economic channel remains protection from punishment, not an increase in the
standalone one-period spot-market deviation gain. None of the results below
changes.
\end{remark}

% User comment: Find language more formal than "destroys" and tone down aggressive phrasing in this part of the theory.
% Fix: Renamed the subsection and proposition around incentive tightening while retaining the existing labels for cross-references.
\subsection{Incentive tightening by protected deviation}
\label{subsec:theory_destabilization}

Define the \emph{punishment slack} of a candidate collusive profile by
\begin{equation}
\label{eq:slack}
S
:=
\delta\bigl(W^C-W^D\bigr)-(\pi^D-\pi^C).
\end{equation}
Without the order book, the collusive-phase incentive constraint is \(S\ge 0\). The next proposition shows that the order book consumes exactly \(B(q,\lambda,L)\) units of that slack.

\begin{proposition}[Incentive tightening by protected deviation]
\label{prop:destabilization}
Under the order-book mechanism, a necessary condition for sustaining the candidate collusive profile is
\begin{equation}
\label{eq:ob_ic_general}
\pi^D-\pi^C+B(q^\star,\lambda,L)
\le
\delta\bigl(W^C-W^D\bigr).
\end{equation}
Equivalently,
\begin{equation}
\label{eq:ob_ic_slack}
S \ge B(q^\star,\lambda,L).
\end{equation}
In the finite-punishment representation, \eqref{eq:ob_ic_general} becomes
\begin{equation}
\label{eq:ob_ic_finite}
\pi^D-\pi^C+B(q^\star,\lambda,L)
\le
\delta\sum_{s=0}^{T-1}\delta^s\bigl(\pi^C-\pi_s^P\bigr).
\end{equation}
In particular, if a collusive profile is sustainable without the order book and satisfies
\[
0\le S < B(q^\star,\lambda,L),
\]
then the order book renders that profile unsustainable.
\end{proposition}

\begin{proof}
With the order book, deviation today yields
\[
V^{D,OB}
=
\pi^D+\delta W^{D,OB}(q^\star,\lambda,L)
=
\pi^D+\delta W^D+B(q^\star,\lambda,L),
\]
where the second equality uses \eqref{eq:defB}. Compliance still yields \(V^C=\pi^C+\delta W^C\). Sequential rationality therefore requires
\[
\pi^C+\delta W^C
\ge
\pi^D+\delta W^D+B(q^\star,\lambda,L),
\]
which rearranges to \eqref{eq:ob_ic_general}. Equation \eqref{eq:ob_ic_slack} is just a rewriting using \eqref{eq:slack}. The finite-punishment form \eqref{eq:ob_ic_finite} follows immediately from Corollary \ref{cor:finite_punishment}. The last claim is immediate.
\end{proof}

\begin{corollary}[Monotone shrinkage of the enforcement region]
	\label{cor:enforcement_region}
	Fix a candidate collusive profile and its best qualifying undercut \(q^\star\).
	Assume that for every punishment period \(s=0,\dots,T-1\) with $L \leq T$ we have
	\(q^\star-c \ge \pi_s^P\).
	Then the set
	\[
	\mathcal A
	:=
	\bigl\{(\lambda,L): S\ge B(q^\star,\lambda,L)\bigr\}
	\]
	of mechanism parameters for which the profile remains sustainable is weakly
	decreasing in both \(\lambda\) and \(L\).  Any parameter change that strictly
	raises \(B\) (which occurs if the inequality is strict in at least one
	protected period) weakly shrinks the set of sustainable collusive profiles.
\end{corollary}

\begin{proof}
	Proposition~\ref{prop:destabilization} shows that a profile is sustainable only
	if \(S \ge B(q^\star,\lambda,L)\).  Under the stated per‑period condition,
	Lemma~\ref{lem:protection}(ii) and (iii) imply that \(B\) is weakly increasing
	in \(\lambda\) and \(L\).  Hence, for a fixed \(S\), the inequality
	\(S \ge B(q^\star,\lambda,L)\) becomes harder to satisfy when \(\lambda\) or
	\(L\) is increased, i.e., \(\mathcal A\) is weakly decreasing in both
	parameters.  If the increase in \(\lambda\) or \(L\) makes at least one
	protected period strictly better for the deviator, then \(B\) strictly
	increases, which (weakly) shrinks \(\mathcal A\) further.
\end{proof}

% User comment: Check elsewhere in the theory section for overly aggressive language.
% Fix: Rephrased the corollary title from "Elimination" to the more formal "Nonexistence" language.
\begin{corollary}[Nonexistence of supracompetitive equilibria within the admissible class]
\label{cor:elimination}
Let \(\mathcal E^{SC}\) denote the set of admissible symmetric public perfect equilibria with collusive price \(p^C>p^N\). For each \(e\in\mathcal E^{SC}\), let \(S(e)\) denote its punishment slack and let \(B_e\) denote its protected deviation value. Define
\[
\bar S := \sup_{e\in\mathcal E^{SC}} S(e),
\qquad
\underline B := \inf_{e\in\mathcal E^{SC}} B_e.
\]
If
\[
\underline B > \bar S,
\]
then no supracompetitive equilibrium in \(\mathcal E^{SC}\) survives under the order-book mechanism.
\end{corollary}

\begin{proof}
Take any \(e\in\mathcal E^{SC}\). Proposition \ref{prop:destabilization} requires \(S(e)\ge B_e\). But \(S(e)\le \bar S < \underline B \le B_e\), a contradiction. Hence no \(e\in\mathcal E^{SC}\) survives.
\end{proof}

\begin{proposition}[Critical patience rises under protection]
\label{prop:critical_delta}
Fix a candidate collusive profile with finite punishment path, and define
\[
F_0(\delta)
:=
\delta\sum_{s=0}^{T-1}\delta^s\bigl(\pi^C-\pi_s^P\bigr)-(\pi^D-\pi^C),
\]
and
\[
F_{OB}(\delta)
:=
F_0(\delta)
-
\delta\sum_{s=0}^{L-1}\delta^s\Bigl(\pi_s^{P,OB}(q^\star,\lambda)-\pi_s^P\Bigr).
\]
Suppose \(F_0\) and \(F_{OB}\) are continuous and strictly increasing on \((0,1)\), and each has a unique root, denoted by \(\delta_0\) and \(\delta_{OB}\) respectively. If
\[
B(q^\star,\lambda,L;\delta_0)>0,
\]
then
\[
\delta_{OB}>\delta_0.
\]
\end{proposition}

\begin{proof}

At the baseline threshold \(\delta_0\), one has \(F_0(\delta_0)=0\). Therefore

\[
F_{OB}(\delta_0)
=
F_0(\delta_0)-B(q^\star,\lambda,L;\delta_0)
=
-B(q^\star,\lambda,L;\delta_0)
<
0.
\]
Since \(F_{OB}\) is strictly increasing and crosses zero uniquely at \(\delta_{OB}\), that root must lie strictly to the right of \(\delta_0\).
\end{proof}

\subsection{Buyers and a weak self-limiting property}
\label{subsec:theory_buyers}

Suppose a buyer may pre-commit a share \(\theta\in(0,1]\) of demand for the next \(L\) periods at price \(q\). Let \(P_{t+s}\) denote the random spot price in period \(t+s\), and let \(\beta\in(0,1]\) be the buyer's discount factor. Assume that commitment affects expenditure only; the utility from consuming the good and the quality of the good supplied are unchanged.

Define the buyer's expected present-value gain from commitment relative to remaining in the spot market by
\begin{equation}
\label{eq:buyer_gain}
G_t(q,\theta)
:=
\theta\sum_{s=0}^{L-1}\beta^s\bigl(\mathbb E_t[P_{t+s}]-q\bigr).
\end{equation}
A nonnegative value of \(G_t(q,\theta)\) means that commitment is weakly preferable.

\begin{proposition}[Buyer rationality and weak self-limiting property]
\label{prop:buyers}
For any \(\theta>0\), a buyer weakly prefers commitment at price \(q\) if and only if
\begin{equation}
\label{eq:buyer_threshold}
q
\le
\frac{\sum_{s=0}^{L-1}\beta^s\mathbb E_t[P_{t+s}]}
     {\sum_{s=0}^{L-1}\beta^s}.
\end{equation}
Consequently:
\begin{enumerate}
\item[(i)] if buyers expect supracompetitive spot prices over the commitment horizon and are offered a qualifying contract price below the discounted expected average spot price in \eqref{eq:buyer_threshold}, they weakly prefer to commit;
\item[(ii)] if buyers expect competitive spot prices bounded above by \(p^N\), then any contract with \(q>p^N\) is unattractive;
\item[(iii)] the operator cannot create a supracompetitive commitment price by fiat when buyers expect competitive spot prices. The mechanism is naturally most relevant when there is a supracompetitive price umbrella to undercut.
\end{enumerate}
\end{proposition}

\begin{proof}
Because commitment operates through expected expenditure, the buyer weakly prefers commitment if and only if \(G_t(q,\theta)\ge 0\). Since \(\theta>0\), this holds if and only if
\[
\sum_{s=0}^{L-1}\beta^s\bigl(\mathbb E_t[P_{t+s}]-q\bigr)\ge 0,
\]
which is equivalent to \eqref{eq:buyer_threshold}. Claims (i) and (ii) are immediate special cases. Claim (iii) follows because any contract that buyers are willing to accept in a competitive environment must be priced at or below the competitive benchmark; hence the mechanism has no direct scope to induce buyers to lock themselves into a new supracompetitive contract price.
\end{proof}

\subsection{Extension to asymmetric costs}
\label{subsec:theory_asymmetric}

The same logic extends seller by seller when firms are asymmetric.

\begin{proposition}[Seller-specific incentive constraints under asymmetric costs]
\label{prop:asymmetric}
Suppose seller \(i\) has marginal cost \(c_i\), and let \(\pi_i^C\), \(\pi_i^D\), \(W_i^C\), \(W_i^D\), and \(B_i\) denote seller \(i\)'s collusive profit, one-shot deviation profit, continuation values, and protected deviation value respectively. Then seller \(i\)'s collusive-phase incentive constraint under the order-book mechanism is
\begin{equation}
\label{eq:asymmetric_ic}
\pi_i^D-\pi_i^C+B_i
\le
\delta\bigl(W_i^C-W_i^D\bigr).
\end{equation}
If \eqref{eq:asymmetric_ic} fails for either seller, the candidate collusive profile cannot be sustained.
\end{proposition}

\begin{proof}
The proof is identical to the proofs of Lemma \ref{lem:cv} and Proposition \ref{prop:destabilization}, applied seller by seller. If either seller finds deviation profitable, the proposed profile fails sequential rationality.
\end{proof}

\subsection{Remarks for interpretation}
\label{subsec:theory_remarks}

\begin{remark}[Blindness to the institution]

The theory represents the order book as a pure payoff-side intervention. The qualification rule and protected-demand allocation change realized profits. The players' information sets remain unchanged. This mirrors the computational design in which order-book variables are excluded from the agents' observation space. The comparative static therefore works through payoff consequences, not through a change in information or feasible actions.

\end{remark}

\begin{remark}[Why simulation comparative statics may be nonmonotone]
Lemma \ref{lem:protection} and Corollary \ref{cor:enforcement_region} are \emph{ceteris paribus} incentive results: they hold for a fixed candidate collusive profile, a fixed qualifying deviation, and a fixed punishment path. In a learning model, changing \(\lambda\) or \(L\) also changes how often the order book is activated, which states are visited during training, and how the policy retrains around the mechanism. For that reason, the direct protection-value comparative statics in the theory may fail to map one-for-one into monotone learned outcomes in simulation.
\end{remark}

\begin{remark}[What the theory abstracts from]
The exact simulation uses a trailing reference price, a discrete grid-step qualification rule, and a blind environment-side change in demand allocation and realized payoffs. Those details are intentionally abstracted away here into the stylized objects \(r\), \(\kappa\), \(\lambda\), and \(L\). The theorem section isolates the economic core: the qualifying discount gives the deviator access to protected committed demand, which softens the continuation-value loss that normally enforces supracompetitive pricing.
\end{remark}

\begin{remark}[Input-cost risk and indexed order books]
	\label{rem:indexed_contracts}
	
	The fixed-cost model normalizes input-cost risk away by fixing marginal cost over the life of the order book. In applications with volatile input costs, a seller may be reluctant to accept a fixed nominal contract price \(q\), since the protected segment would expose it to adverse input-price movements. This is an implementation issue rather than a change in the mechanism's enforcement logic.

	One natural implementation is an indexed order-book contract. Let \(X_t\) be a public input-cost index and let a contract accepted at date \(t\) specify
	\[
	q^{OB}_{t+s}=\alpha X_{t+s}+\eta,
	\qquad s=0,\ldots,L-1,
	\]
	where \(\alpha\) is a pass-through coefficient and \(\eta\) is the contractual markup over the index. Qualification can then be defined in markup space rather than in nominal-price space, for example
	\[
	\eta \leq \bar{\eta}_t-\kappa_{\eta},
	\]
	where \(\bar{\eta}_t\) is a recent reference markup implied by spot prices and the same input-cost index. In retail gasoline, \(X_t\) could be a public crude-oil, wholesale gasoline, futures, or local rack-price index. The point is not that the contract must reference any particular index, but that it can target the supracompetitive markup while passing through input-cost movements that are outside the collusive enforcement problem.
	
	Under such an implementation, the protected-profit term in \eqref{eq:protected_profit} is replaced by
	\[
	\pi_s^{P,OB,idx}
	=
	(1-\lambda)\pi_s^P
	+
	\lambda \mathbb{E}_t\!\left[q^{OB}_{t+s}-c_{t+s}\right],
	\]
	or by the realized indexed margin when the contract settles period by period. The protection-value logic is unchanged after replacing \(q-c\) by the relevant expected or realized indexed margin. The same incentive-tightening result applies whenever the indexed contract gives the deviator a higher protected-period payoff than the punishment payoff it would otherwise receive.
\end{remark}

Taken together, Lemma \ref{lem:cv} and Proposition \ref{prop:destabilization} formalize the paper's central theory claim: the order-book mechanism weakens collusive enforcement by cushioning the deviator against future punishment, thereby making qualifying undercuts more attractive. Proposition \ref{prop:buyers} adds the complementary buyer-side point that the institution is naturally self-limiting: it is useful precisely when prevailing prices are expected to remain above competitive benchmarks.

Table \ref{tab:summary_design} summarizes the environment, learner, and mechanism objects used in the paper.

\begin{table}[t]
\centering
\caption{Environment, learner, and mechanism summary}
\label{tab:summary_design}
\begin{tabular}{lp{0.64\textwidth}}
\toprule
Object & Specification in the main design \\
\midrule
Environment & Repeated differentiated-products Bertrand pricing on a finite grid \\
Reference prices & Symmetric Nash price 1.4729; monopoly price 1.9250 \\
Learner & Strategic DQN with the same architecture in both environments \\
State variables & Own price level, rival price level, rival movement type, rival movement duration \\
State count & 375 strategic states \\
Action set & Deep cut, qualifying cut, tactical undercut, match, soft follow, strong follow \\
Mechanism & Blind temporary customer-commitment order book \\
Main order-book parameters & Share 0.50, duration 5, ten-period reference, three-step qualification \\
Main evaluation & Separate baseline environment and environment with the mechanism, trained from scratch \\
Evaluation object & Final-policy greedy rollout realized transaction price \\
\bottomrule
\end{tabular}
\end{table}

\section{Testing the Order-Book Mechanism}
\label{sec:experiment}

\subsection{Evaluation design}
\label{subsec:evaluation_design}

The main evaluation is a paired comparison in which both environments are trained from scratch. The baseline environment disables the order-book mechanism. The environment with the mechanism activates the same mechanism from period 0. Both environments use the same strategic DQN architecture, state representation, action set, training horizon, and exploration schedule. The market institution is the only design difference; learner information and algorithm are held fixed.

This design matches an operator-led market-design intervention. In a platform implementation, the marketplace platform changes the institutional design of the market directly. In an independent-aggregator implementation, the same demand block can enter the market from outside the incumbent selling platform. In either case, pricing systems are likely to be retrained, retuned, or otherwise updated under the new environment. The policy-relevant outcome is the deployed pricing policy learned under the new institution. The from-scratch design measures that outcome directly.

The main evaluation uses final-policy greedy rollouts. In each paired run, the policies learned in the baseline environment and the environment with the mechanism are evaluated using the same rollout protocol. The outcome is the mean realized transaction price. In the environment with the mechanism, realized transaction price incorporates both spot-market purchases and any committed-demand transactions generated by the order book. This consumer-facing price measure captures the mechanism's consequences for posted prices, demand allocation, and realized payoffs. Appendix \ref{app:implementation} gives the exact rollout construction, including the reference-start warmup and the common-state diagnostic starts.

The comparison has a second layer aimed at interpretation rather than only measuring average price differences. After measuring the price difference between the environments with and without the mechanism, the paper returns to the trained checkpoints and compares them from common states. These additional diagnostics include high-start greedy rollouts from identical high-price histories and same-state forced-deviation exercises in which one seller is forced to make a qualifying cut while the rival is held fixed. Because the mechanism is blind to the agents' observation and action spaces, these tests can be read as evidence on how the institution changes deviation incentives, rather than as consequences of added information or a different control problem.

\subsection{Main symmetric-cost results}
\label{subsec:main_results}

% User comment: Skip paired direction-count reporting in the paper.
% Fix: Removed paired direction-count totals and kept the inference focused on means, price differences between the environments with and without the mechanism, and confidence intervals.

\begin{figure}[t]
\centering
\includegraphics[width=0.82\textwidth]{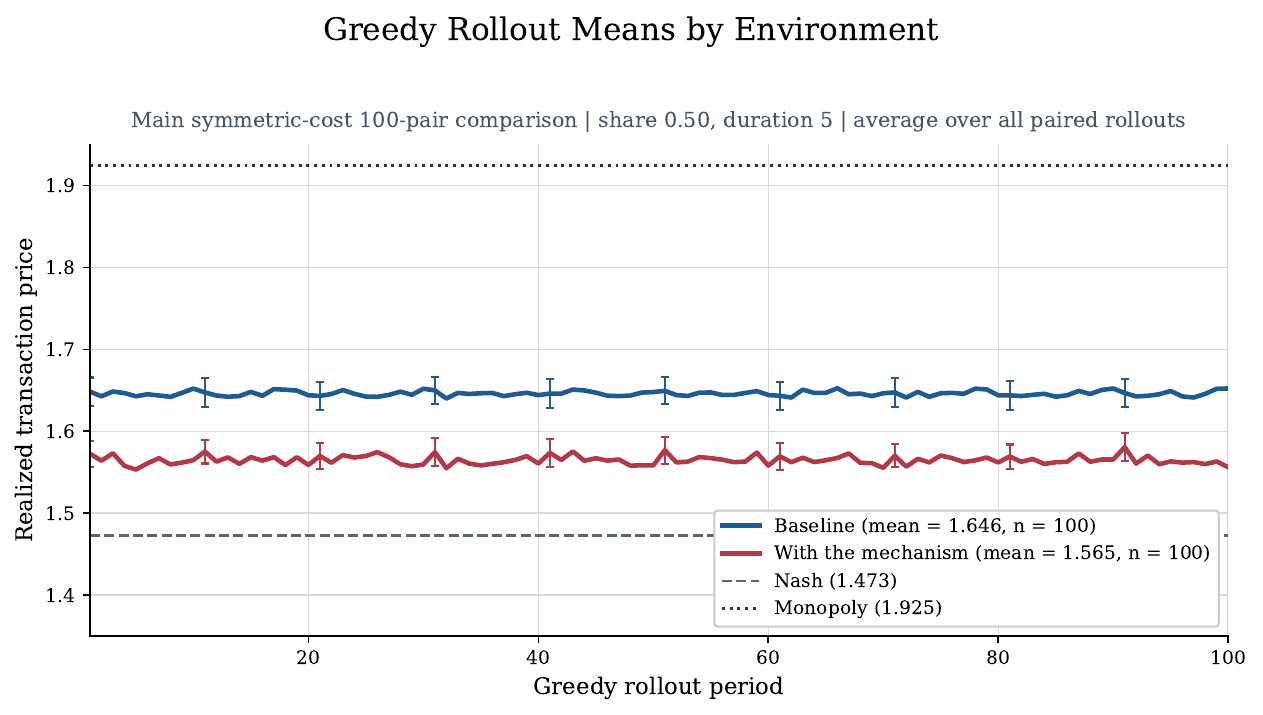}
\caption{Main symmetric-cost greedy rollout comparison. Both environments are trained from scratch. The environment with the mechanism has a 0.50 committed-demand share, five-period duration, ten-period reference window, and three-step qualifying discount.}
\label{fig:main_with_without}
\end{figure}

Figure \ref{fig:main_with_without} is the headline result. In the symmetric-cost 100-pair comparison, the mean realized greedy-rollout price is 1.6461 in the baseline environment and 1.5647 in the environment with the mechanism. The average price difference between the environment with the mechanism and the baseline environment is -0.0814. The approximate 95 percent confidence interval is [-0.101, -0.061].

The normalized price difference is economically substantial. Relative to the Nash price of 1.4729, the policy learned in the baseline environment is about 11.75 percent above Nash and the policy learned in the environment with the mechanism is about 6.23 percent above Nash. Equivalently, the mechanism closes about 47.0 percent of the baseline collusion gap. This is the paper's main quantitative result: retraining under the order-book institution materially lowers final-policy realized prices in the main symmetric-cost design.

Table \ref{tab:main_results} summarizes the main quantitative results across the symmetric-cost design and the heterogeneous-cost extension. The table reports final-policy greedy-rollout outcomes. For the symmetric row, the percent-over-Nash calculation uses the symmetric Nash price of 1.4729. For the heterogeneous-cost row, it uses the heterogeneous-cost realized Nash price of 1.5656.

\begin{table}[t]
\centering
\caption{Main quantitative results}
\label{tab:main_results}
\small
\setlength{\tabcolsep}{4pt}
\begin{tabular}{lrrrrrr}
\toprule
Design & Baseline & With the mechanism & Diff. & \% over Nash & Gap red. & 95\% CI \\
\midrule
Symmetric cost & 1.6461 & 1.5647 & -0.0814 & 11.75 / 6.23 & 47.0\% & [-0.101, -0.061] \\
Heterogeneous cost & 1.6920 & 1.6469 & -0.0450 & 8.07 / 5.19 & 35.6\% & [-0.068, -0.022] \\
\bottomrule
\end{tabular}
\begin{minipage}{0.92\textwidth}
\footnotesize
\textit{Notes:} ``Diff.'' is the environment with the mechanism minus the baseline environment. ``\% over Nash'' reports the baseline value followed by the value with the mechanism. Gap reduction is measured relative to the relevant Nash price.
\end{minipage}
\end{table}

The confidence interval reinforces the point. The interval for the main symmetric-cost design is well below zero. The result is an average final-policy price difference across paired runs. The paper's central claim is that the order-book institution lowers learned pricing outcomes on average.

\subsection{Mechanism validation: targeting the enforcement channel}
\label{subsec:diagnostics}

A lower rollout mean in the environment with the mechanism is not by itself a mechanism result, since deep neural networks are often difficult to interpret: an intervention can appear successful while the economic channel remains opaque. The present framework permits a sharper interpretation because the order-book institution is blind to the agents' observation and action spaces. Any price difference between the environments with and without the mechanism must therefore arise from changed payoff consequences, not from additional information or a richer control problem.

A first diagnostic is that active-book periods are too rare to explain the main result mechanically. In the symmetric-cost evaluation, only 10 of 100 rollouts with the mechanism contain any active-book periods. Mean active-book share in the environment with the mechanism is about 5.7 percent and mean book-take-up share is about 1.2 percent. Yet Figure \ref{fig:main_with_without} shows a substantial price difference between the environments with and without the mechanism. This pattern suggests that the institution changes the learned pricing policy itself, rather than merely discounting a small number of realized transactions on the deployment path.

The most direct test is a same-state forced-deviation decomposition. Starting from common high-price histories, the hold path keeps both sellers at the high start price at \(t=0\). The deviation path holds the rival fixed at that same price but forces the focal seller to make a qualifying three-step cut. From \(t=1\) onward, both agents are released back to greedy play. For the deviator, the diagnostic decomposes the discounted value change from the forced cut into the period-0 payoff component, the continuation loss from later punishment, and the net deviation value, defined as deviation-path value minus hold-path value. This separates the ordinary one-period price-cutting payoff from the dynamic punishment channel. The exercise is repeated for 100 paired runs, four high start prices, and both possible deviators, giving \(100\times 4\times 2=800\) forced deviations.

\begin{table}[t]
\centering
\caption{Forced-deviation mechanism diagnostics}
\label{tab:forced_deviation}
\begin{tabular}{lrrrr}
\toprule
Diagnostic & Baseline & With the mechanism & Diff. & 95\% CI \\
\midrule
Period-0 payoff component & 0.0422 & 0.1529 & 0.1108 & [0.1091, 0.1125] \\
Continuation loss & 0.2883 & -0.4140 & -0.7023 & [-0.7654, -0.6393] \\
Net deviation value & -0.2462 & 0.5669 & 0.8131 & [0.7498, 0.8764] \\
Order-book take-up at \(t=0\) & 0.0000 & 1.0000 & 1.0000 & [1.0000, 1.0000] \\
\bottomrule
\end{tabular}
\begin{minipage}{0.92\textwidth}
\footnotesize
\textit{Notes:} Entries are means over 800 forced deviations: 100 paired runs, four high-price starts, and two possible deviators. ``Diff.'' is the environment with the mechanism minus the baseline environment. The confidence interval is for the paired difference. Order-book take-up is an indicator that the forced deviator receives the book at \(t=0\).
\end{minipage}
\end{table}

Table \ref{tab:forced_deviation} shows that the results line up closely with the intended channel. The mechanism reduces the continuation-loss component and raises net deviation value. In environment-level means, the same qualifying cut is unprofitable in the baseline environment but profitable in the environment with the mechanism. The continuation term flips sign as well, from a positive continuation loss in the baseline environment to a continuation gain in the environment with the mechanism. In the forced-deviation tests with the mechanism, the deviator takes up the order book at \(t=0\) in all cases, compared with no take-up in the baseline environment. Because same-period take-up can be recorded in the period-0 component, that component should be interpreted as the first accounting period of protection rather than as a larger standalone spot-market deviation payoff. The sign pattern is broad-based: the continuation-loss difference is negative in 610 of 800 deviations and in 95 of 100 run-level averages, and the net-deviation-value difference is positive in 642 of 800 deviations and in 97 of 100 run-level averages.

A second dynamic diagnostic starts both environments from the same high-price histories and then releases them to greedy play. In these high-start rollouts, the policies learned in the environment with the mechanism end the path about 0.1295 lower in the last-20-period mean spot-price window. The path with the mechanism is lower in 338 of 400 high-start tests and in 89 of 100 run-level averages. This is consistent with the deviation-value evidence: when protected deviations face smaller continuation penalties, high-price states are less persistent under the policy learned in the environment with the mechanism.

The Q-advantage diagnostic is supportive but secondary. Feeding common one-hot strategic states through each trained network shows that in high-price hold states the networks trained in the environment with the mechanism value cut actions more highly than networks trained in the baseline environment. The average cut-action advantage is higher by about \(+0.0391\), and the greedy action is a cut in 46.9 percent of high-hold cases in the environment with the mechanism versus 18.4 percent in the baseline environment.

Taken together, the diagnostics imply more than a reduced-form price reduction. The mechanism lowers prices for the designed economic reason. Qualifying undercuts are partially protected from subsequent punishment, making continuation losses materially smaller and leading the learned policy to place less weight on sustaining high-price states.

\subsection{Heterogeneous-cost extension}
\label{subsec:heterogeneous}

% User comment: Mention in the heterogeneous-cost discussion that this extension differentiates the paper from Johnson's symmetric-firm setting.
% Fix: Added the comparison with Johnson while keeping the subsection focused on the heterogeneous-cost result.
The heterogeneous-cost extension tests the mechanism in an unequal-cost environment. In this design, seller costs are 1.0 and 1.2, and the known heterogeneous-cost Nash prices are approximately 1.5330 and 1.6100. The realized Nash price is about 1.5656. The order-book parameters are otherwise the same as in the main design: a 0.50 committed-demand share, five-period duration, ten-period reference window, and three-step qualification threshold. This extension is useful for comparison with \citet{johnson2020platform}, whose closest market-design benchmark focuses on symmetric firms; the order-book evidence here includes asymmetric sellers.

\begin{figure}[t]
\centering
\includegraphics[width=0.82\textwidth]{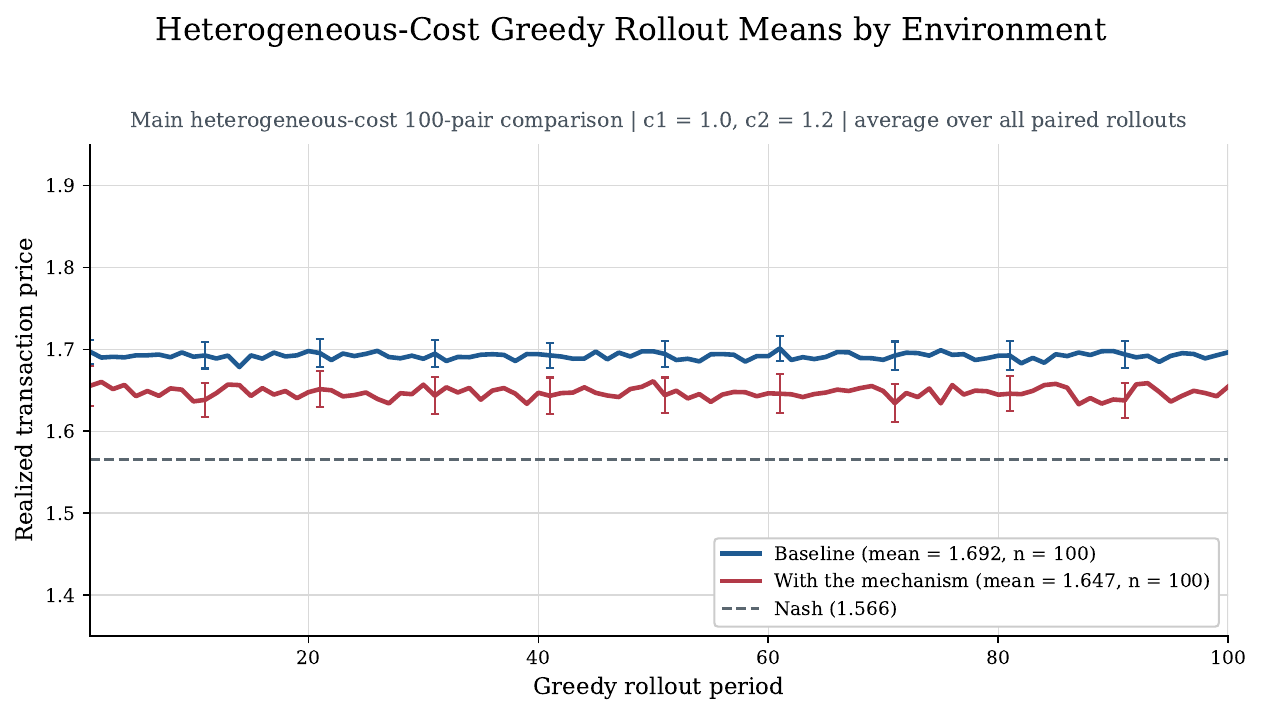}
\caption{Heterogeneous-cost greedy rollout comparison. Seller costs are 1.0 and 1.2. The mechanism remains effective on average. The price difference is smaller than in the main symmetric-cost design.}
\label{fig:heterogeneous}
\end{figure}

Figure \ref{fig:heterogeneous} reports the 100-pair heterogeneous-cost comparison. The mean realized rollout price is 1.6920 in the baseline environment and 1.6469 in the environment with the mechanism. The price difference between the environment with the mechanism and the baseline environment is -0.0450, with an approximate 95 percent confidence interval of [-0.068, -0.022]. Relative to the heterogeneous-cost realized Nash price, the policy learned in the baseline environment is about 8.07 percent above Nash and the policy learned in the environment with the mechanism is about 5.19 percent above Nash. The baseline heterogeneous-cost collusion gap is \(1.6920-1.5656=0.1264\), and the mechanism closes that gap by \(0.0450/0.1264\approx 35.6\) percent.

This extension shows that the order-book mechanism is effective in markets with heterogeneous costs (practically, heterogeneous qualities) as well. This differentiates it from the DPDP mechanism in \citet{johnson2020platform}, which is developed and analyzed under symmetric constant marginal costs.\footnote{Dynamic PDP does not extend straightforwardly to heterogeneous-cost settings. Its logic is based on raw price comparisons: the seller with the lowest price receives the advantage, and keeps it unless it raises price or is undercut by more than a common cushion. That is natural when sellers have the same marginal cost, because a lower price can be interpreted as a more aggressive competitive move. With heterogeneous costs, however, the lower-cost seller will typically post a lower price even under ordinary competitive play, so the mechanism would partly reward cost advantage itself rather than a deviation from collusive pricing. In a simple two-cost case, a five-cent cut may be minor for the low-cost seller but substantial for the high-cost seller, so a common threshold no longer has a common meaning. A heterogeneous-cost version would therefore require seller-specific or cost-adjusted thresholds, such as comparisons in markups or deviations from seller-specific benchmark prices, which in turn requires additional information and changes the mechanism's incentive logic.}

% User comment: Note that collusion is less significant in the heterogeneous-cost extension, giving the mechanism less space to act, and provide collusion-gap percentages.
% Fix: Added the heterogeneous-cost collusion-gap arithmetic and explained the smaller price difference relative to the smaller baseline gap.

The heterogeneous-cost result is an economically important extension. The average price difference between the environments with and without the mechanism remains negative away from exact cost symmetry. The price difference is smaller than in the symmetric-cost design partly because the baseline collusion gap is smaller: 0.1264 in the heterogeneous-cost extension, compared with 0.1732 in the symmetric-cost benchmark. The mechanism closes about 35.6 percent of the heterogeneous-cost gap, leaving the policy learned in the environment with the mechanism about 5.19 percent above the realized Nash price. The price difference remains negative and economically meaningful.

\subsection{Additional robustness and parameter sweep}
\label{subsec:robustness}

A neutral-start rollout check evaluates both environments from the same shared midpoint state. The qualitative conclusion survives: the final policy learned in the environment with the mechanism remains lower than the final policy learned in the baseline environment. This check supports the interpretation that the main result reflects learned policy behavior, with little role for differences in the reference histories used to initialize rollouts.

The parameter sweep is also informative. Low book shares do little and can be near neutral in this environment. The mechanism becomes strongest in the higher-share, multi-period region, especially around shares 0.40 to 0.50 with duration above one period. Appendix Table \ref{tab:sweep} reports the sweep compactly, and Appendix Figure \ref{fig:sweep_appendix} reproduces the share-specific figure panels.

% User comment: Use "Literature Review" and avoid "positioning" language.
% Fix: Retitled the section and renamed the subsections so the literature review reads as a standard academic section.
\section{Literature Review}
\label{sec:literature}

\subsection{Strategic complexity and bounded-memory representations}
\label{subsec:bounded_memory_lit}

The strategic-state framework is connected to an older repeated-games literature on bounded memory and strategic complexity. \citet{rubinstein1986finiteautomata} introduced finite automata as a way to model repeated-game strategies with limited complexity. \citet{abreu1988finiteautomata} study Nash equilibria in repeated games when players' preferences depend both on repeated-game payoffs and on the complexity of the automata implementing their strategies. \citet{kalai1988finite} define strategy complexity by the number of distinct continuation strategies induced across subgames and show that finite-complexity strategies can approximate repeated-game equilibrium payoffs.

A closely related bounded-memory literature studies repeated-game behavior when strategies depend only on a finite window or compressed representation of past play. \citet{sabourian1998boundedmemory} characterizes repeated games with \(M\)-period bounded memory, while \citet{barlo2009onememory} and \citet{barlo2016boundedmemory} show that even severe memory restrictions can support rich repeated-game incentives under appropriate conditions. This literature is motivated partly by limits on human memory and cognition, but its logic is useful for algorithmic-pricing experiments as well. The relevant takeaway is not that any particular finite state is sufficient. Rather, it is that the representation of history is a substantive modeling choice: a compact state can preserve strategically relevant information while avoiding the curse of dimensionality associated with raw histories.

The present paper uses that idea operationally. It maps the recent public price history into a finite strategic state consisting of own price level, rival price level, rival movement direction, and movement duration. This is not claimed to be an optimal sufficient statistic for repeated pricing. Instead, it is a theory-guided state abstraction whose adequacy is assessed behaviorally. The rise/drop validation asks whether the learned policy uses the compressed state in an economically sensible way: punishing rival price cuts while accommodating rival price increases.

\subsection{Q-learning collusion and market design}
\label{subsec:qlearning_lit}

This paper builds on the foundational Q-learning collusion literature. \citet{calvano2020artificial} show that independent pricing algorithms can learn supracompetitive outcomes in a repeated Bertrand environment. \citet{klein2021autonomous} studies autonomous Q-learning under sequential pricing. These papers make algorithmic collusion a concrete object of economic analysis. The present paper builds from that foundation toward intervention design, using a compact strategic DQN framework with explicit directional behavioral validation.

A related methodological issue is whether supracompetitive prices reflect a strategically meaningful collusive policy or a learning artifact. \citet{calvano2023algorithmic} formalize this distinction by separating genuine collusion from spurious collusion driven by exploration and learning dynamics. \citet{banchio2022artificial} provide a different mechanism, spontaneous coupling, in which learning algorithms become statistically linked in a way that supports supracompetitive outcomes. These contributions reinforce the role of the behavioral validation in Section \ref{subsec:behavioral_validation}: the DQN policy learned in the baseline environment is not interpreted as collusive merely because average prices are high, but because it punishes rival cuts and accommodates rival price increases.

\citet{arunachaleswaran2025algorithmic} study another distinct channel: algorithmic collusion without explicit threats. Their main result concerns sequential algorithm deployment, where a first mover uses a no-regret learner and a second mover optimizes in the induced environment. That setting is different from the simultaneous DQN adoption studied here. The DQN learners in the present paper are value-based reinforcement-learning agents with function approximation and do not carry a formal no-regret guarantee. The comparison is therefore useful precisely because it separates two mechanisms: no-regret algorithmic commitment on the one hand, and the continuation-value enforcement channel validated and targeted here on the other.

Finally, the algorithmic-monoculture literature highlights a separate competition concern: firms may use similar models, shared data sources, common benchmarks, or correlated prediction systems, producing market outcomes that are more aligned than the firms' formal independence would suggest \citep{kleinberg2021algorithmic,jo2025homogeneous}. This channel is distinct from the punishment channel studied here. The present paper holds the learner class fixed and uses behavioral and forced-deviation diagnostics to show how a payoff-side institution changes the incentives around undercutting.

% User comment: Clarify the comparison with Johnson by discussing DPDP, its 70 percent market allocation, homogeneous-cost scope, and the behavioral ambiguity of the Calvano benchmark highlighted by the Economics Letters paper.
% Fix: Expanded the Johnson paragraph with those details while keeping the comparison constructive.
\citet{johnson2020platform} is the closest market-design predecessor. Their Dynamic Price Directed Prominence (DPDP) mechanism rewards a seller that obtains an advantage by routing a large share of demand toward that seller; in the simulations emphasized in that paper, the advantaged seller receives 70 percent of the market. That rule shows clearly how platform allocation can alter collusive incentives. It also raises a practical capacity and fulfillment issue: a seller that receives most of the market may be unable or unwilling to serve that volume in product markets with supply constraints. The DPDP mechanism is also developed for homogeneous-cost sellers, where a lower posted price has a common strategic interpretation across firms. With heterogeneous costs, a raw price advantage can partly reflect cost asymmetry instead of a deviation from a collusive arrangement. While DPDP operates through platform demand steering, the present order-book design uses temporary customer commitment mediated by an operator that may be either the platform or an independent profit-seeking aggregator. Section \ref{subsec:heterogeneous} evaluates the order-book mechanism with unequal seller costs.

The simulation environment also differs. The computational evidence in \citet{johnson2020platform} builds on the Calvano-style Q-learning environment. While that environment is foundational, \citet{epivent2024rewardpunishment} show that the trained learner can punish upward deviations as well as downward deviations. Such behavior is difficult to interpret as a theoretically disciplined collusive response because a rival price increase should normally be accommodated rather than punished. The DQN framework used here therefore adds the rise/drop validation in Section \ref{subsec:behavioral_validation}: rival cuts are followed by lower prices, while rival increases are generally accommodated. The intervention is then tested in a testbed where the trained algorithm better aligns with the directional comparative statics of standard collusive enforcement.

The broader repeated-game and collusion literature provides the economic language used in Section \ref{sec:model}. Classic theories of collusive enforcement emphasize continuation-value losses, punishment, and the sustainability of supracompetitive prices \citep{stigler1964theory,green1984noncooperative,1988Abreu}. The order-book mechanism can be read as a market-design attempt to alter those enforcement constraints. The design targets the payoff consequences of price cutting through observable market rules.

\subsection{Deep-learning collusion}
\label{subsec:deep_lit}

% User comment: Give more detail on the deep-learning collusion papers and make the distinction the behavioral test of theoretically sound responses.
% Fix: Rewrote the subsection around the main deep-RL comparators and made the rise/drop validation the explicit differentiating feature.

\citet{hettich2021algorithmic} is the closest predecessor on DQN and state compression. They replace tabular Q-learning with DQN in a Calvano-style repeated-pricing environment, and show that deep learning can speed convergence to collusive outcomes. They also study how collusion changes with the number of firms, and propose a state reformulation based on the average, minimum, and maximum of last-period prices. The paper also verifies reward-punishment behavior by forcing a unilateral cut to the static Nash price and observing retaliation. That validation establishes punishment after a downward deviation. The present framework adds a complementary test by separating downward and upward rival moves to test economic understanding among the trained agents: a crucial step for the evaluation of any policy intervention.

\citet{schlechtinger2023price} and \citet{ijcai2024p54} study deep reinforcement learning pricing agents in richer market simulations. Their paper uses DQN and PPO, allows flexible demand conditions, varies the number of sellers and demand bias, and studies a restricted-information case in which agents cannot observe rivals' prices. Their action design is relative to the agent's own previous price, and the results document robust supracompetitive outcomes, including oscillatory pricing patterns. \citet{deng2024algorithmic} provide a broad algorithm comparison across tabular Q-learning, DQN, PPO, and SAC in several Bertrand-type environments, including standard Bertrand, Bertrand-Edgeworth, and logit-demand Bertrand markets. Their results show that collusion depends on both the algorithm and the market model, with market structure and algorithm class changing the strength and form of supracompetitive pricing.

\citet{friedrich2025learning} study DQN and PPO in episodic, inventory-constrained markets, where finite selling horizons and limited inventory change the repeated-game logic. They compute competitive and collusive benchmarks numerically, study forced deviations and best-response surfaces, and show learned strategic structure, including lower pricing after opponent deviations and a ``climb the hill'' dynamic toward collusion.\footnote{This does not include tests about responses to opponents raising prices.} \citet{han2025relative} study independent learners with relative experience replay and show that relative-performance concerns affect collusive outcomes: agents more tolerant of underperformance tend toward supracompetitive prices, while agents more averse to underperformance tend toward Bertrand-Nash. \citet{kitamura2026influence} focus on state representation and report that, in a three-firm setting, a representation based on the minimum and second-minimum prices can generate stronger collusion than conventional state summaries. \citet{frick2026convergence} studies continuous-price soft actor-critic learning and shows that modern policy-gradient methods can reach supracompetitive outcomes on much shorter training horizons than standard Q-learning. The paper validates learned collusion by forcing a one-period static-best-response deviation and by visualizing deterministic policy dynamics through phase diagrams. These diagnostics show punishment and return to high-price play, such as seen in \citet{calvano2020artificial}.

The concern also extends beyond standard product-market pricing. \citet{dou2025aipowered} study AI-powered trading and show that reinforcement-learning traders can sustain collusive profits in financial markets, with consequences for price efficiency. Their setting is different from the repeat-purchase product markets studied here, but it underscores the broader point that autonomous learning systems can create competition problems across market institutions.

These papers establish that modern deep reinforcement learning can generate supracompetitive pricing across several algorithms, market structures, and state representations. The distinctive framework feature here is the behavioral validation exercise. The learner in the baseline environment is tested with rise and drop perturbations and delivers the economically disciplined response required for a collusive-policy interpretation: it lowers price after rival cuts and generally raises price after rival increases. That validation is important for the policy intervention exercise because a market-design intervention is most informative when the baseline environment is both supracompetitive and strategically interpretable.

\subsection{Policy interventions and audits}
\label{subsec:remedies_audits}

Aside from \citet{johnson2020platform} (discussed above), \citet{brero2022learning} is the closest intervention paper. They study a computational platform intervention. Their platform uses Stackelberg reinforcement learning to learn buy-box allocation rules that mitigate collusion by reinforcement-learning sellers. The platform is the leader, sellers are learning followers, and the mechanism works by changing the allocation rule that sellers face. \citet{brero2022learning} are therefore an important bridge between platform-design interventions and machine-learning implementation. Their seller side is built around Q-learning agents, while the present paper evaluates a market-design intervention against strategic DQN pricing agents with the behavioral validation described above.

Recent work also studies interventions that operate directly through the learning rule. \citet{chica2024algorithmic} analyze Q-learning collusion in two-sided markets with network externalities and propose a penalty-term intervention that reduces collusive outcomes. That approach changes the algorithmic objective. The order-book mechanism studied here instead changes the market institution: it alters realized allocation and payoffs while leaving the pricing agents' state representation and action space unchanged.

% User comment: Explain the Hartline appendix exercise in more detail using the local HartlineStage2 materials.
% Fix: Added the noncollusive Nash benchmark, the 5,000- and 10,000-round audit numbers, and the reason the exercise is used as a false-positive benchmark.

\citet{hartline2024regulation} and \citet{hartline2025refined} study audit and certification approaches. Their framework asks whether logged pricing behavior can be certified as plausibly noncollusive using regret-type conditions, including pessimistic calibrated regret in the refined version. This is a different regulatory margin from the one studied here. Audit rules are ex post informational tools: they evaluate behavior after prices have been generated. The order-book mechanism is an ex ante market-design tool: it changes the payoff environment in which pricing policies are learned.

Appendix \ref{app:hartline} reports a small calibration exercise that illustrates this distinction. In the numerical environment used in this paper, a deliberately noncollusive stage-game learner converges to prices near the Bertrand-Nash price. When a refined Hartline-style certificate is applied to the resulting transcript, certification is not obtained at the sample sizes considered. This illustrates that certification can be conservative in finite samples, and that audit-based regulation and incentive-based market design are complementary rather than substitutes.

% User comment: Add that the order-book intervention is market-driven, can be run by an aggregator outside the platform, and can be privately profitable, distinguishing it from audits and Johnson's platform-dependent DPDP.
% Fix: Added a paragraph on independent aggregation and the profit/implementation distinction.
The order-book institution also differs in implementation. The mechanism is the temporary order book; the operator is the entity that first collects conditional buyer commitments, thereby assembles an order book, and then offers that book to sellers. A qualifying seller can buy the book and fill the committed demand block. One implementation is platform-operated: the selling platform itself acts as the buyer aggregator. A second implementation is independent aggregation: an outside intermediary operates the order book and earns a small spread or fee between the buyer commitment price and the supplier contract price. This makes the intervention a market-driven institution. Audit regimes require legal or regulatory authority to impose certification requirements, and the DPDP requires the platform to alter its demand-routing rule. The order-book mechanism can in principle be run by an intermediary that profits from organizing temporary customer commitments.

\section{Conclusion}
\label{sec:conclusion}

This paper develops an interpretable framework for studying learned collusion and uses it to evaluate whether market design can weaken collusion by deep-learning pricing algorithms. It introduces an order-book mechanism under which an operator first assembles temporary buyer commitments and then offers the resulting order book to sellers willing to supply at sufficiently low prices. The operator may be the selling platform or an independent buyer-side aggregator. The mechanism is blind to the agents' observations and operates through realized demand allocation and payoffs. In the strategic DQN framework with supracompetitive pricing in the baseline environment, retraining with the mechanism lowers final-policy realized prices from 1.6461 to 1.5647 in the main symmetric-cost design, closing about 47 percent of the collusion gap.

The paper's main claim is interpretive as well as quantitative. The baseline environment exhibits the economically sensible asymmetry of punishing cuts and accommodating rises. Most importantly, same-state forced-deviation diagnostics show that the mechanism works through the intended channel: qualifying undercuts become materially less vulnerable to future punishment because the order book protects part of the deviator's demand during the punishment phase.

The larger implication is that market design deserves a more central place in the policy discussion of algorithmic collusion. Detection and audit tools are important, but institutional design can also make collusion harder to sustain. The evidence here supports that institutional approach and shows that the evaluation of a policy intervention is most informative when paired with transparent mechanism validation.

\appendix
\setcounter{figure}{0}
\setcounter{table}{0}
\renewcommand{\thefigure}{A\arabic{figure}}
\renewcommand{\thetable}{A\arabic{table}}
\section{Implementation Details}
\subsection{Implementation Details of the Strategic DQN Simulator}
\label{app:implementation}

This appendix records the implementation-level specification of the main computational environment. The training, evaluation, and diagnostic routines all implement the same economic environment, state encoder, action map, order-book rule, and DQN update. The large paired runs simply train many independent seller pairs in parallel; this changes the computational layout, not the economic model.

A minimal implementation can follow the following structure. Initialize a price grid and an eight-period history for each seller pair. At each period, encode each seller's history into a 375-state one-hot observation, let each seller choose one of six relative-price actions, convert those actions into posted price indices, apply the order-book allocation rule if the mechanism is active, compute profits, append the new price pair to the history, and store each seller's transition \((s_t,a_t,r_t,s_{t+1})\). The DQN update then samples stored transitions, updates the policy network toward a Double-DQN target, and periodically copies policy-network weights into the target network. Evaluation uses the same environment and state encoder but turns off exploration.

\subsubsection{Primitive objects and price grid}

There are two sellers, indexed by \(i\in\{1,2\}\), with constant marginal cost \(c=1.0\). Demand is Bertrand-logit with parameters \(A=2.0\), \(A_0=0.0\), and \(\mu=0.25\). The discrete price grid has 15 equally spaced values between 1.00 and 2.10. Let the grid step be
\[
\Delta=\frac{2.10-1.00}{14}\approx 0.07857.
\]
If \(k\in\{0,\dots,14\}\) is a price index, the corresponding posted price is
\[
p(k)=1.00+k\Delta.
\]
The simulator stores the last eight price-index pairs, so the history at decision time \(t\) is an \(8\times 2\) array
\[
H_t=\bigl[(k_{t-7,1},k_{t-7,2}),\dots,(k_{t,1},k_{t,2})\bigr].
\]
All state variables observed by the agents are deterministic functions of this eight-period index history. The order-book institution never enters the observation directly.

\subsubsection{State space, observation vector, and output space}

For agent \(i\), the strategic state has four components.

First, the simulator coarse-codes the agent's own current price index and the rival's current price index into five price-level bins by integer division by three:
\[
\ell(k)=\min\{\lfloor k/3\rfloor,4\}.
\]
Thus indices \(0\)--\(2\) map to \texttt{LL}, \(3\)--\(5\) to \texttt{L}, \(6\)--\(8\) to \texttt{M}, \(9\)--\(11\) to \texttt{H}, and \(12\)--\(14\) to \texttt{HH}.

Second, the simulator constructs the rival's recent movement regime from the rival's eight-period price-index history. Let the rival's history be \(v_0,\dots,v_7\), and define first differences \(d_\tau=v_\tau-v_{\tau-1}\) for \(\tau=1,\dots,7\). The code finds the last nonzero difference, takes its sign as the current regime direction, and then scans backward. Earlier zero differences are skipped, earlier nonzero differences with the same sign are accumulated into the regime's total change, and the scan stops at the first nonzero difference with the opposite sign. This produces a signed total change \(g\) and a regime start date.

Third, the total signed change \(g\) is bucketed into five rival-movement categories:
\[
\texttt{sharp\_drop}\ \text{if}\ g\le -3,\qquad
\texttt{mild\_drop}\ \text{if}\ g\in\{-2,-1\},
\]
\[
\texttt{hold}\ \text{if}\ g=0,\qquad
\texttt{mild\_rise}\ \text{if}\ g\in\{1,2\},\qquad
\texttt{sharp\_rise}\ \text{if}\ g\ge 3.
\]
The cutoffs are in price-grid steps, not in price levels.

Fourth, the regime duration is converted into three buckets. If the current regime has lasted at most two periods, the duration bucket is \texttt{fleeting}; if it has lasted three to four periods, it is \texttt{brief}; otherwise it is \texttt{sustained}. In the code, a completely flat rival history is encoded as a hold regime with duration equal to the full eight-period history, so a long unchanged price is encoded as \texttt{hold} plus \texttt{sustained}.

The total state space is therefore
\[
5\times 5\times 5\times 3=375
\]
states. If \(o_i\) is own price level, \(r_i\) is rival price level, \(m_i\) is rival movement category, and \(d_i\) is rival duration bucket, the state identifier used by the code is
\[
s_i=\bigl(5o_i+r_i\bigr)\cdot 15+\bigl(3m_i+d_i\bigr)\in\{0,\dots,374\}.
\]
The neural network input is the one-hot vector \(e_{s_i}\in\{0,1\}^{375}\).

The network output space is six-dimensional. For each state, the policy network and the target network each return a vector in \(\mathbb{R}^6\), one \(Q\)-value for each strategic action. Greedy play chooses the index of the largest component. During training, epsilon-greedy exploration instead replaces the greedy choice with a random action with probability \(\varepsilon_t\).

\subsubsection{Action map}

Each action is defined as an offset from the rival's current price index, not from the agent's own current price. Let \(k_{t,j}\) be the rival's current price index at the end of the history window. If agent \(i\) chooses action \(a\), the next posted index is
\[
k'_{t+1,i}=\min\bigl\{14,\max\{0,k_{t,j}+\omega(a)\}\bigr\},
\]
where the action offsets are
\[
\omega(a)\in\{-4,-3,-1,0,1,3\}.
\]
The named actions are:
\begin{itemize}
\item \texttt{deep\_cut}: \(-4\);
\item \texttt{qualifying\_cut}: \(-3\);
\item \texttt{tactical\_undercut}: \(-1\);
\item \texttt{match}: \(0\);
\item \texttt{soft\_follow}: \(+1\);
\item \texttt{strong\_follow}: \(+3\).
\end{itemize}
The boundary clamp matters in low-price and high-price states. For example, a deep cut from a rival index of 1 is clipped to index 0, and a strong follow from a rival index of 13 is clipped to index 14.

\subsubsection{Within-period timing and payoff rule}

The simulator does not explicitly model a separate buyer-solicitation stage.
Economically, when the order-book institution is active, the period begins with
a menu of contingent buyer-side commitments indexed by qualifying contract
prices. The code implements this menu in reduced form: after sellers choose
candidate supply prices, the environment checks whether any candidate supply
price takes up one of the pre-assembled contingent books.

The simulator's period-\(t\) timing is as follows.

\begin{enumerate}
\item The environment begins with an eight-period history \(H_t\) and, possibly, an inherited active order book from earlier periods.
\item Each agent observes its one-hot 375-state vector computed from \(H_t\).
\item Each agent chooses one of the six relative-price actions. During training the choice is epsilon-greedy; during evaluation it is greedy.
\item The environment converts the chosen actions into candidate supply-price
indices \(k'_{t+1,1}\) and \(k'_{t+1,2}\), then into candidate supply prices
\(p_1\) and \(p_2\).
\item If the order-book institution is active, the simulator computes a reference price
\[
r_t=\frac{1}{W_t}\sum_{\tau=t-W_t+1}^{t}\bar p_\tau,
\]
where \(\bar p_\tau\) is the mean posted spot price in period \(\tau\) and
\(W_t=\min\{10,\text{available prior periods}\}\). The reference price is
computed from prior spot-price means only; it does not include the current
candidate supply prices being evaluated.
\item The minimum qualifying contract price is
\[
q_t^{\min}=r_t-3\Delta.
\]
Economically, buyers have already indicated willingness to commit a share
\(\lambda=0.50\) of demand for five periods at qualifying contract prices.
\item The agents' chosen prices are then interpreted as candidate supply prices.
A seller takes up the pre-assembled order book if its candidate supply price is
weakly below \(q_t^{\min}\). If no book is already active and several sellers
qualify, the seller with the lowest qualifying price takes up the book. Ties are
broken randomly. The contract price is the winner's current candidate supply
price, frozen for the life of the contract.
\item A newly activated book lasts for five periods including the take-up period. While a book is active, no new book can be taken up.
\item Spot demand is always computed from the current posted prices using standard logit shares
\[
s_i(p_1,p_2)=\frac{\exp((A-p_i)/\mu)}{\exp(A_0/\mu)+\exp((A-p_1)/\mu)+\exp((A-p_2)/\mu)}.
\]
Without an active book, seller \(i\)'s profit is
\[
\pi_i=(p_i-c)s_i.
\]
\item With an active book of share \(\lambda=0.50\) owned by seller \(b\), the code carves the market into a residual spot segment and a committed segment. Residual spot demand is
\[
D_i^{\text{spot}}=(1-\lambda)s_i.
\]
Committed demand is
\[
D_i^{\text{book}}=
\begin{cases}
\lambda & \text{if } i=b,\\
0 & \text{otherwise.}
\end{cases}
\]
If the book owner's contract price is \(\hat p_b\), profits become
\[
\pi_b=(p_b-c)(1-\lambda)s_b+(\hat p_b-c)\lambda,
\]
\[
\pi_j=(p_j-c)(1-\lambda)s_j\qquad (j\neq b).
\]
The realized transaction price reported in the paper is the quantity-weighted average across residual spot transactions at current posted prices and committed transactions at the frozen contract price.
\item The environment appends the new price pair to the history, appends the current mean posted price to the trailing reference-price buffer, decrements the book's remaining life by one, and carries the book forward if at least one protected period remains.
\end{enumerate}

This timing implies that a current-period take-up can be recorded as protection beginning in the take-up period itself. That accounting convention should not be interpreted as a separate increase in the standalone one-period spot-market deviation payoff. The mechanism operates through realized allocation and payoffs by protecting committed demand from later punishment. No order-book state variable enters the agents' observation vectors.

\subsubsection{Training algorithm}

Each seller has its own DQN, target network, optimizer state, replay buffer, and epsilon counter. In the main large-sample implementation, the 100 paired environments are trained in parallel on one device, but weights are not shared across environments or across agents within an environment. The parallelization is purely computational.

For each training step \(t\), the implementation performs the following operations:
\begin{enumerate}
\item compute the two sellers' state identifiers from the current eight-period history;
\item choose actions using epsilon-greedy behavior, independently for each seller;
\item map actions into next posted price indices and apply the payoff rule described above;
\item compute the next state identifiers after appending the new price pair to the history;
\item store \((s_t,a_t,r_t,s_{t+1})\) in the seller's replay buffer; and
\item if enough transitions have been accumulated, sample a minibatch and apply the Double-DQN update below.
\end{enumerate}

The policy network has architecture
\[
375 \rightarrow 128 \rightarrow 128 \rightarrow 6
\]
with ReLU activations after the two hidden layers. The target network has the same architecture. The optimizer is Adam with learning rate \(10^{-4}\). The replay capacity is 100,000 transitions per agent, the minibatch size is 64, the discount factor is \(\gamma=0.95\), and the loss is Huber loss.

For each sampled transition \((s_t,a_t,r_t,s_{t+1})\), the code uses a Double-DQN target:
\[
y_t=r_t+\gamma Q^{-}\!\left(s_{t+1},\arg\max_a Q(s_{t+1},a)\right),
\]

where \(Q\) is the policy network and \(Q^{-}\) is the target network. The reward \(r_t\) is simply the seller's own current-period profit. There is no terminal state in the Bellman target; the task is modeled as continuing. The target network is hard-updated every 1,000 steps. Gradients are clipped elementwise at 1.0 in the large-sample parallel implementation.

Exploration is epsilon-greedy with raw schedule
\[
\varepsilon_t=0.01+(0.90-0.01)\exp(-t/250000).
\]

The environment is reset every 250 periods. A reset reinitializes the eight-period history and clears any active book, but it does not reset the learned network weights, the replay buffers, or the epsilon counters. The main symmetric-cost evaluation uses 100 paired seeds trained for 1,000,000 steps in each environment. In the environment with the mechanism, the order-book institution is active from period 0; in the baseline environment, the same code is used but the order-book institution is disabled.

\subsubsection{Evaluation protocol and diagnostic starts}

The paper's main outcome is the mean realized transaction price in a final-policy greedy rollout. The standard reference-start evaluation runs the trained checkpoint greedily for 100 periods. To construct the starting history, the evaluation procedure first takes a 200-period no-book greedy rollout from the symmetric midpoint history, that is, from eight periods in which both sellers begin at the midpoint grid index \(7\). The final eight-period window of that warmup path becomes the rollout start history. This protocol gives each checkpoint an internally consistent starting state that is close to its learned attractor.

The neutral-start robustness check instead begins directly from the common midpoint history with no warmup. This produces the same state for both environments by construction.

The mechanism diagnostics use additional common starts. High-start rollouts begin from histories in which both sellers have repeated the same high index for all eight periods; the implementation uses indices 9, 10, 11, and 12. These high-start tests run for 80 periods. The forced-deviation decomposition uses the same high starts but a 60-period horizon. In period 0, one seller is forced to cut by exactly three grid steps while the rival is held at the start price. From period 1 onward both sellers return to greedy play. This is the implementation-level diagnostic behind the period-0 payoff, continuation-loss, and net-deviation-value decomposition reported in the main text.

Finally, the heterogeneous-cost extension keeps the same state encoder, action map, network architecture, and evaluation logic, while changing the seller costs to 1.0 and 1.2.

\subsection{Parameter Sweep}
\label{app:sweep}

Table \ref{tab:sweep} reports the share-by-duration sweep compactly. Each cell gives the mean realized rollout price for the corresponding order-book parameter pair, evaluated against the same baseline symmetric-cost mean of 1.6461.

\begin{table}[h]
\centering
\caption{Parameter sweep summary}
\label{tab:sweep}
\begin{tabular}{lrrrrp{0.28\textwidth}}
\toprule
Share & Duration 1 & Duration 3 & Duration 5 & Duration 8 & Summary \\
\midrule
0.10 & 1.6657 & 1.6767 & 1.6709 & 1.6580 & Weak or adverse \\
0.20 & 1.6761 & 1.6691 & 1.6741 & 1.6527 & Weak or adverse \\
0.30 & 1.6723 & 1.6361 & 1.6405 & 1.6449 & Near neutral \\
0.40 & 1.6608 & 1.5709 & 1.6176 & 1.5928 & Effective when multi-period \\
0.50 & 1.6558 & 1.5692 & 1.5647 & 1.5696 & Strongest region \\
\bottomrule
\end{tabular}
\end{table}

The sweep also points to a deployment caveat. Cases in which the environment with the mechanism has a higher price than the baseline environment appear only in low-participation regions and are small in magnitude; none survives Holm correction across the 20 sweep configurations. In the tested grid, the region with committed-buyer share at least 0.30 and commitment duration at least 5 exhibits no evidence of adverse price increases. Thus, the mechanism should be viewed as a participation-contingent intervention: it is most appropriate once sufficient committed demand has been assembled, rather than as a tool to deploy unconditionally.

\begin{figure}[p]
\centering
\begin{subfigure}{0.48\textwidth}
  \centering
  \includegraphics[width=\textwidth]{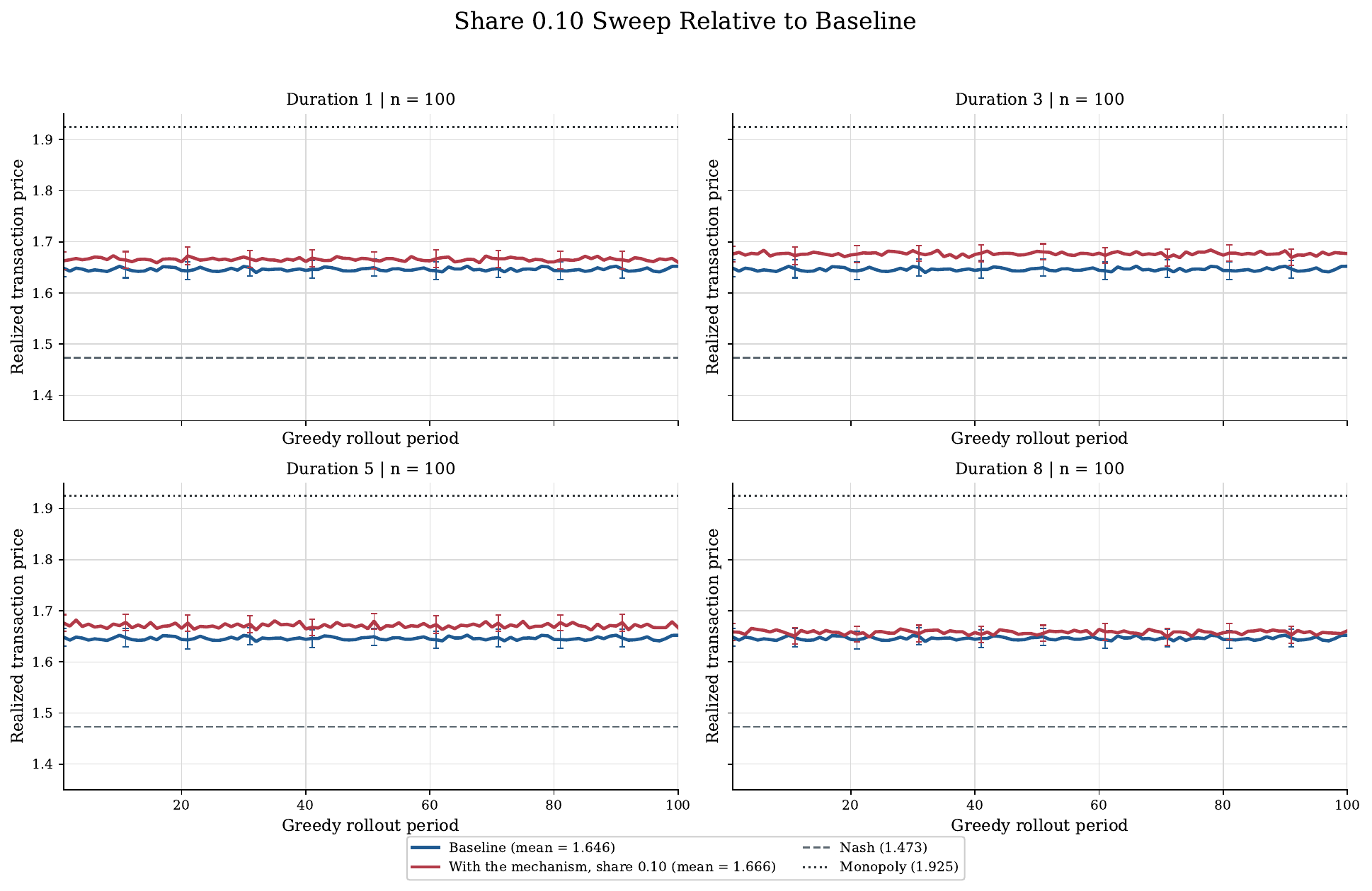}
  \caption{Share 0.10}
\end{subfigure}
\hfill
\begin{subfigure}{0.48\textwidth}
  \centering
  \includegraphics[width=\textwidth]{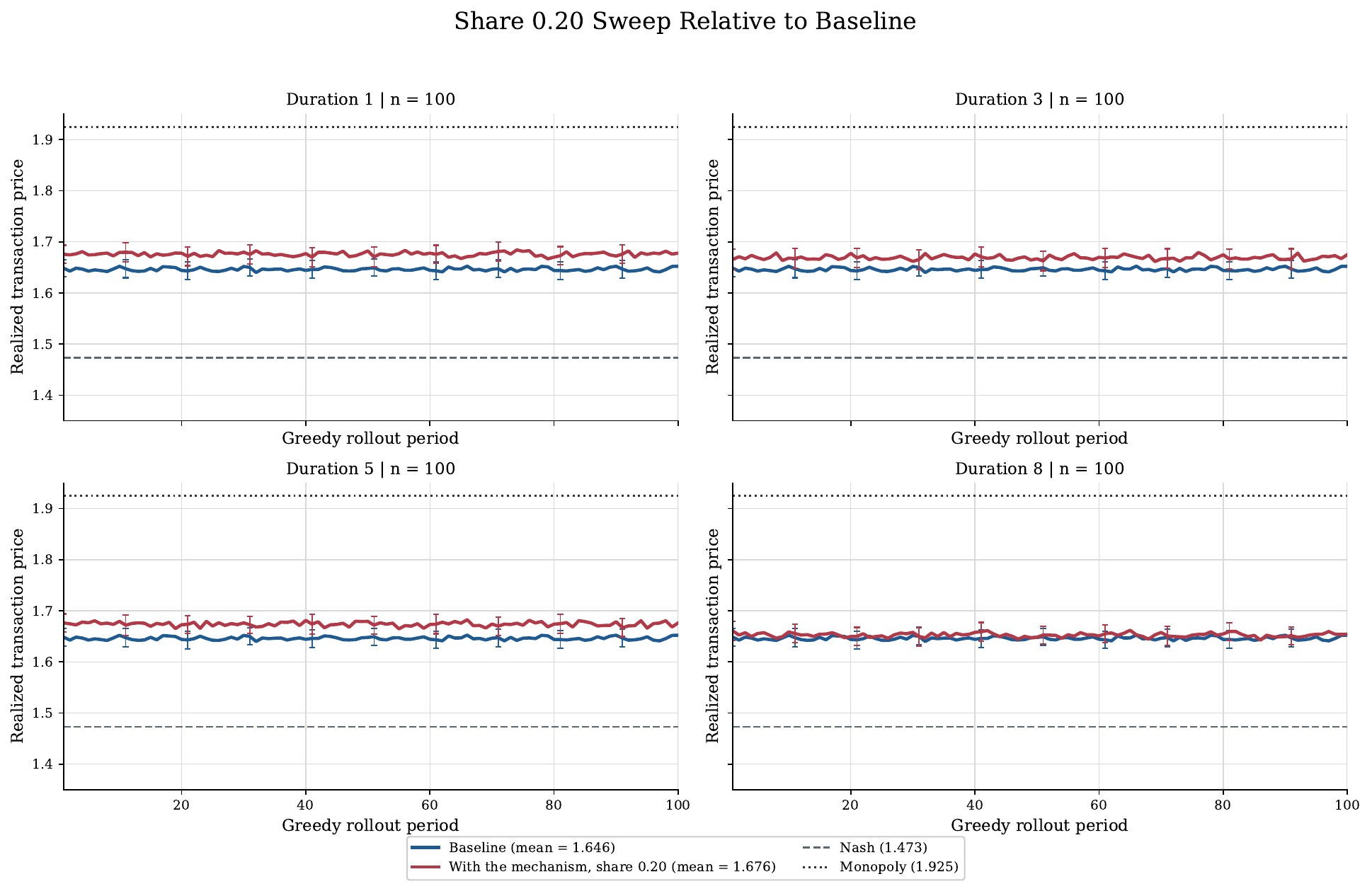}
  \caption{Share 0.20}
\end{subfigure}

\vspace{0.5em}
\begin{subfigure}{0.48\textwidth}
  \centering
  \includegraphics[width=\textwidth]{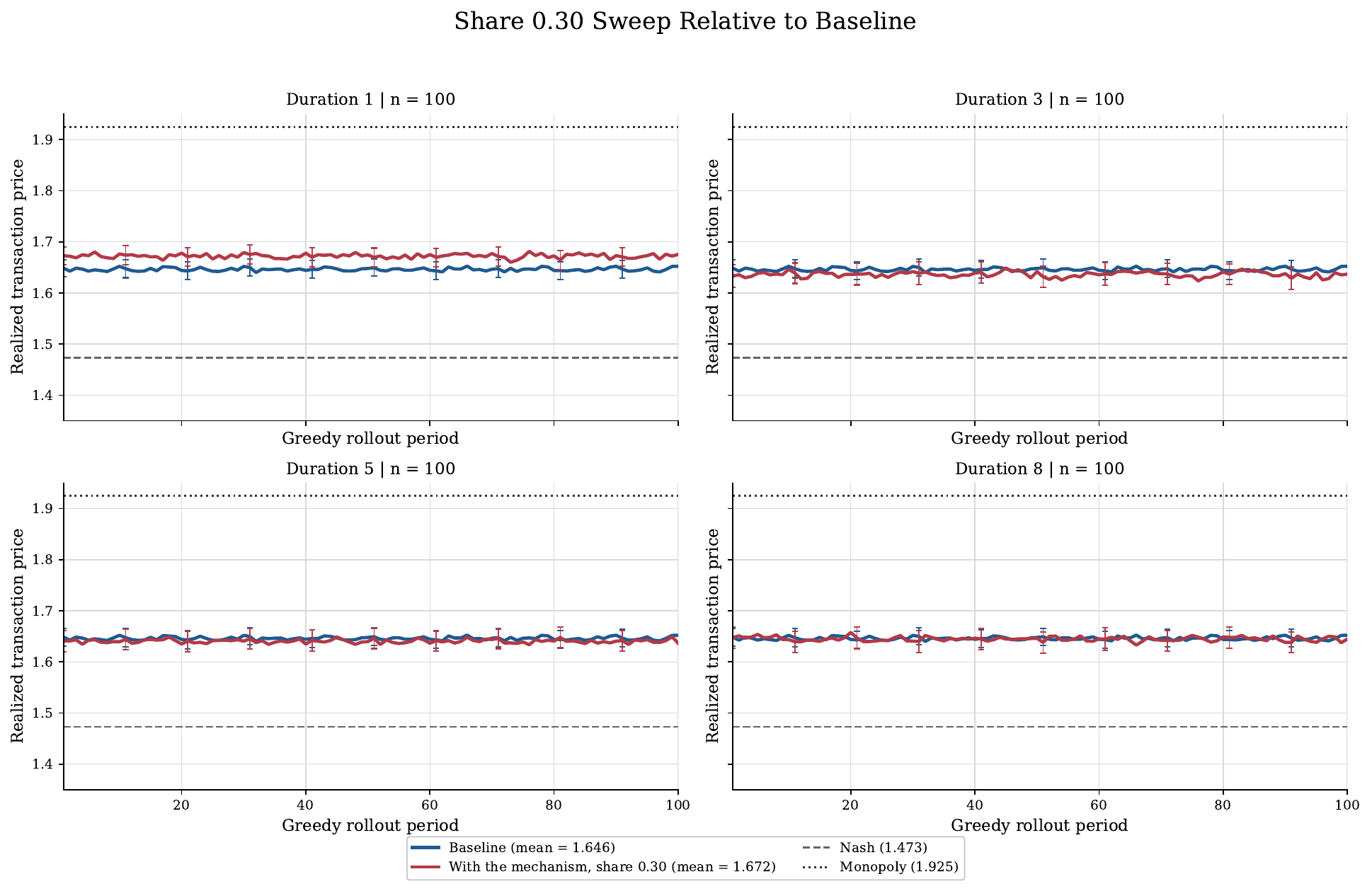}
  \caption{Share 0.30}
\end{subfigure}
\hfill
\begin{subfigure}{0.48\textwidth}
  \centering
  \includegraphics[width=\textwidth]{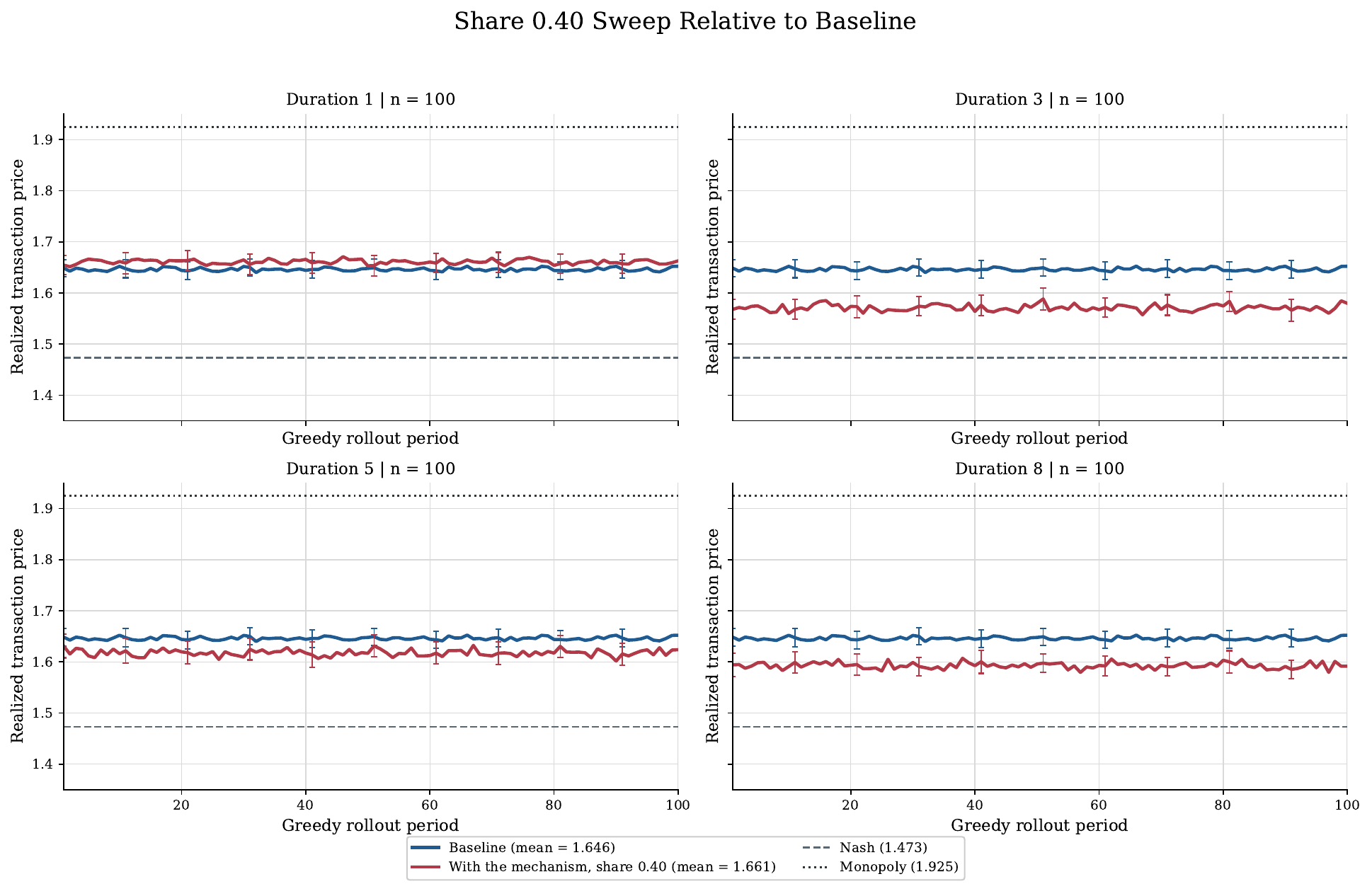}
  \caption{Share 0.40}
\end{subfigure}

\vspace{0.5em}
\begin{subfigure}{0.60\textwidth}
  \centering
  \includegraphics[width=\textwidth]{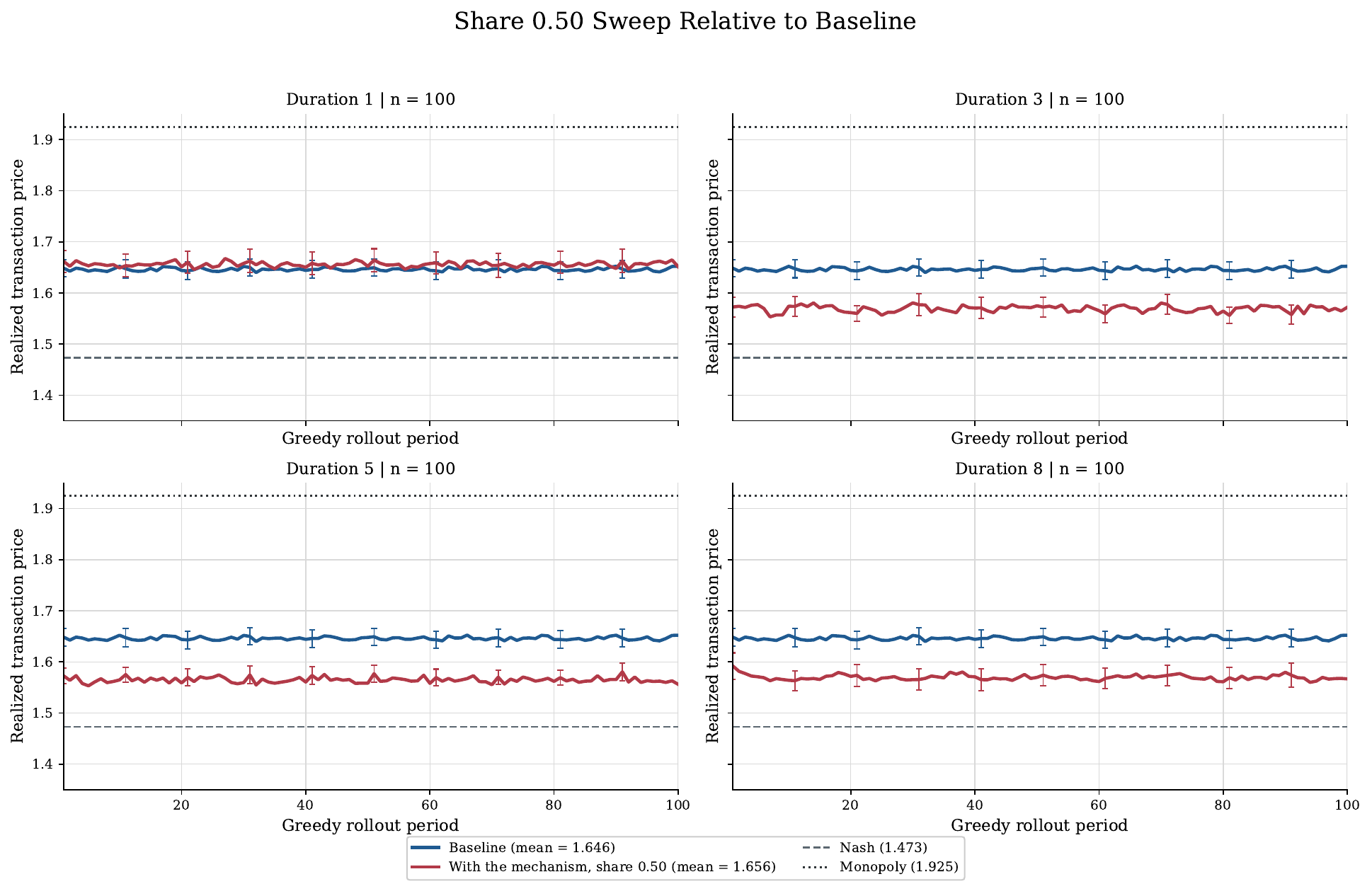}
  \caption{Share 0.50}
\end{subfigure}
\caption{Share-by-duration sweep. The panels compare each parameter cell with the mechanism to the baseline symmetric-cost environment.}
\label{fig:sweep_appendix}
\end{figure}

\subsection{Audit Discussion and Finite-Sample Certification}
\label{app:hartline}

This appendix records a small calibration exercise related to audit-based approaches. \citet{hartline2024regulation} and \citet{hartline2025refined} study certification rules based on regret-type properties. Such tools ask whether logged behavior can be certified as plausibly noncollusive. The order-book mechanism targets a different object: the incentive consequences of market institutions for learned final-policy prices.

The exercise constructs a deliberately noncollusive stage-game learner in the numerical setting shared with \citet{calvano2020artificial}. The learner is based on fictitious play, best responds to the opponent's empirical price distribution, and converges to the symmetric Bertrand-Nash price while maintaining a small probability of playing every action. Its greedy prices are around 1.4714, close to the Nash price.

Applying a refined Hartline-style certificate to this transcript does not deliver certification at the sample sizes considered. At 5,000 rounds, the estimated pessimistic regrets are 0.0559 for seller 1 and 0.0773 for seller 2. The confidence term is about 486.4, so neither seller satisfies a pass threshold of 0.1. At 10,000 rounds, seller 1's estimated pessimistic regret falls to 0.0240, but the confidence term remains about 481.2, and certification is still not obtained.

This exercise is not intended as a general critique of audit frameworks. It is a finite-sample calibration in one numerical environment. Its role is to illustrate that noncertification can arise even for behavior generated by a competitive stage-game learner, depending on the sample size, confidence term, and pass threshold. This reinforces the conceptual distinction between ex post certification and ex ante market design. Audit tools may be valuable for detection and enforcement, while market-design interventions can be valuable because they alter the incentives that generate pricing behavior in the first place.
\pagebreak
\bibliographystyle{apalike}
\bibliography{col}

\begin{thebibliography}{}

\bibitem[Abreu, 1988]{1988Abreu}
Abreu, D. (1988).
\newblock On the theory of infinitely repeated games with discounting.
\newblock {\em Econometrica: Journal of the Econometric Society}, pages 383--396.

\bibitem[Abreu and Rubinstein, 1988]{abreu1988finiteautomata}
Abreu, D. and Rubinstein, A. (1988).
\newblock The structure of {Nash} equilibrium in repeated games with finite automata.
\newblock {\em Econometrica}, 56(6):1259--1281.

\bibitem[Arunachaleswaran et~al., 2025]{arunachaleswaran2025algorithmic}
Arunachaleswaran, E.~R., Collina, N., Kannan, S., Roth, A., and Ziani, J. (2025).
\newblock {Algorithmic Collusion Without Threats}.
\newblock In Meka, R., editor, {\em 16th Innovations in Theoretical Computer Science Conference (ITCS 2025)}, volume 325 of {\em Leibniz International Proceedings in Informatics (LIPIcs)}, pages 10:1--10:21, Dagstuhl, Germany. Schloss Dagstuhl -- Leibniz-Zentrum f{\"u}r Informatik.

\bibitem[Assad et~al., 2024]{clark2023algorithmic}
Assad, S., Clark, R., Ershov, D., and Xu, L. (2024).
\newblock Algorithmic pricing and competition: Empirical evidence from the german retail gasoline market.
\newblock {\em Journal of Political Economy}, 132(3):723--771.

\bibitem[Banchio and Mantegazza, 2022]{banchio2022artificial}
Banchio, M. and Mantegazza, G. (2022).
\newblock Artificial intelligence and spontaneous collusion.
\newblock {\em arXiv preprint arXiv:2202.05946}.
\newblock Last revised September 2023.

\bibitem[Barlo et~al., 2009]{barlo2009onememory}
Barlo, M., Carmona, G., and Sabourian, H. (2009).
\newblock Repeated games with one-memory.
\newblock {\em Journal of Economic Theory}, 144(1):312--336.

\bibitem[Barlo et~al., 2016]{barlo2016boundedmemory}
Barlo, M., Carmona, G., and Sabourian, H. (2016).
\newblock Bounded memory folk theorem.
\newblock {\em Journal of Economic Theory}, 163:728--774.

\bibitem[Brero et~al., 2022]{brero2022learning}
Brero, G., Mibuari, E., Lepore, N., and Parkes, D.~C. (2022).
\newblock Learning to mitigate {AI} collusion on economic platforms.
\newblock In {\em Advances in Neural Information Processing Systems}, volume~35, pages 37892--37904.

\bibitem[Brown and MacKay, 2023]{brown2023competition}
Brown, Z.~Y. and MacKay, A. (2023).
\newblock Competition in pricing algorithms.
\newblock {\em American Economic Journal: Microeconomics}, 15(2):109--156.

\bibitem[Calvano et~al., 2020]{calvano2020artificial}
Calvano, E., Calzolari, G., Denicolo, V., and Pastorello, S. (2020).
\newblock Artificial intelligence, algorithmic pricing, and collusion.
\newblock {\em American Economic Review}, 110(10):3267--97.

\bibitem[Calvano et~al., 2023]{calvano2023algorithmic}
Calvano, E., Calzolari, G., Denicol{\`o}, V., and Pastorello, S. (2023).
\newblock Algorithmic collusion: Genuine or spurious?
\newblock {\em International Journal of Industrial Organization}, 90:102973.

\bibitem[Chica et~al., 2024]{chica2024algorithmic}
Chica, C., Guo, Y., and Lerman, G. (2024).
\newblock Artificial intelligence and algorithmic price collusion in two-sided markets.
\newblock {\em arXiv preprint arXiv:2407.04088}.

\bibitem[Deng et~al., 2024]{deng2024algorithmic}
Deng, S., Schiffer, M., and Bichler, M. (2024).
\newblock Algorithmic collusion in dynamic pricing with deep reinforcement learning.
\newblock {\em arXiv preprint arXiv:2406.02437}.

\bibitem[Dou et~al., 2025]{dou2025aipowered}
Dou, W.~W., Goldstein, I., and Ji, Y. (2025).
\newblock {AI}-powered trading, algorithmic collusion, and price efficiency.
\newblock NBER Working Paper 34054, National Bureau of Economic Research.

\bibitem[Epivent and Lambin, 2024]{epivent2024rewardpunishment}
Epivent, A. and Lambin, X. (2024).
\newblock On algorithmic collusion and reward--punishment schemes.
\newblock {\em Economics Letters}, 237:111661.

\bibitem[Frick, 2026]{frick2026convergence}
Frick, K.~M. (2026).
\newblock Convergence to collusion in algorithmic pricing.
\newblock {\em arXiv preprint arXiv:2604.15825}.

\bibitem[Friedrich et~al., 2025]{friedrich2025learning}
Friedrich, P., Pasztor, B., and Ramponi, G. (2025).
\newblock Learning collusion in episodic, inventory-constrained markets.
\newblock In {\em Proceedings of the 24th International Conference on Autonomous Agents and Multiagent Systems (AAMAS '25)}, pages 803--812, Richland, SC. International Foundation for Autonomous Agents and Multiagent Systems.

\bibitem[Green and Porter, 1984]{green1984noncooperative}
Green, E.~J. and Porter, R.~H. (1984).
\newblock Noncooperative collusion under imperfect price information.
\newblock {\em Econometrica: Journal of the Econometric Society}, pages 87--100.

\bibitem[Han, 2025]{han2025relative}
Han, B. (2025).
\newblock Algorithmic pricing with independent learners and relative experience replay.
\newblock In {\em Proceedings of the 6th ACM International Conference on AI in Finance (ICAIF '25)}, pages 80--87.
\newblock Earlier circulated as arXiv:2102.09139 under the title ``Understanding algorithmic collusion with experience replay''.

\bibitem[Hartline et~al., 2024]{hartline2024regulation}
Hartline, J.~D., Long, S., and Zhang, C. (2024).
\newblock Regulation of algorithmic collusion.
\newblock {\em arXiv preprint arXiv:2401.15794}.

\bibitem[Hartline et~al., 2025]{hartline2025refined}
Hartline, J.~D., Wang, C., and Zhang, C. (2025).
\newblock Regulation of algorithmic collusion, refined: Testing pessimistic calibrated regret.
\newblock {\em arXiv preprint arXiv:2501.09740}.

\bibitem[Hettich, 2021]{hettich2021algorithmic}
Hettich, M. (2021).
\newblock Algorithmic collusion: Insights from deep learning.
\newblock {\em SSRN Electronic Journal}.

\bibitem[Jo et~al., 2025]{jo2025homogeneous}
Jo, N., Creel, K., Wilson, A., and Raghavan, M. (2025).
\newblock Homogeneous algorithms can reduce competition in personalized pricing.
\newblock In {\em Advances in Neural Information Processing Systems}, volume~38.
\newblock NeurIPS 2025.

\bibitem[Johnson et~al., 2023]{johnson2020platform}
Johnson, J.~P., Rhodes, A., and Wildenbeest, M. (2023).
\newblock Platform design when sellers use pricing algorithms.
\newblock {\em Econometrica}, 91(5):1841--1879.

\bibitem[Kalai and Stanford, 1988]{kalai1988finite}
Kalai, E. and Stanford, W. (1988).
\newblock Finite rationality and interpersonal complexity in repeated games.
\newblock {\em Econometrica}, 56(2):397--410.

\bibitem[Kitamura and Fujita, 2026]{kitamura2026influence}
Kitamura, Y. and Fujita, K. (2026).
\newblock Influence of state representation on algorithmic collusion under deep learning.
\newblock In Mathieu, P. and De~la Prieta, F., editors, {\em Advances in Practical Applications of Agents, Multi-Agent Systems, and Computational Social Science: The PAAMS Collection}, volume 16031 of {\em Lecture Notes in Computer Science}, pages 203--215.

\bibitem[Klein, 2021]{klein2021autonomous}
Klein, T. (2021).
\newblock Autonomous algorithmic collusion: Q-learning under sequential pricing.
\newblock {\em The RAND Journal of Economics}, 52(3):538--558.

\bibitem[Kleinberg and Raghavan, 2021]{kleinberg2021algorithmic}
Kleinberg, J. and Raghavan, M. (2021).
\newblock Algorithmic monoculture and social welfare.
\newblock {\em Proceedings of the National Academy of Sciences}, 118(22):e2018340118.

\bibitem[Rubinstein, 1986]{rubinstein1986finiteautomata}
Rubinstein, A. (1986).
\newblock Finite automata play the repeated prisoner's dilemma.
\newblock {\em Journal of Economic Theory}, 39(1):83--96.

\bibitem[Sabourian, 1998]{sabourian1998boundedmemory}
Sabourian, H. (1998).
\newblock Repeated games with {M}-period bounded memory (pure strategies).
\newblock {\em Journal of Mathematical Economics}, 30(1):1--35.

\bibitem[Schlechtinger et~al., 2024]{ijcai2024p54}
Schlechtinger, M., Kosack, D., Krause, F., and Paulheim, H. (2024).
\newblock By fair means or foul: Quantifying collusion in a market simulation with deep reinforcement learning.
\newblock In Larson, K., editor, {\em Proceedings of the Thirty-Third International Joint Conference on Artificial Intelligence, IJCAI-24}, pages 485--493.

\bibitem[Schlechtinger et~al., 2023]{schlechtinger2023price}
Schlechtinger, M., Kosack, D., Paulheim, H., Fetzer, T., and Krause, F. (2023).
\newblock The price of algorithmic pricing: Investigating collusion in a market simulation with ai agents.
\newblock In {\em Proceedings of the 22nd International Conference on Autonomous Agents and Multiagent Systems (AAMAS '23)}, pages 2748--2750.

\bibitem[Stigler, 1964]{stigler1964theory}
Stigler, G.~J. (1964).
\newblock A theory of oligopoly.
\newblock {\em Journal of Political Economy}, 72(1):44--61.

\end{thebibliography}

\end{document}